\documentclass[graybox]{svmult}

\usepackage{makeidx}         

\usepackage{graphicx}        

\usepackage{multicol}        

\usepackage[bottom]{footmisc}

\usepackage{amssymb,amsmath,amsfonts}

\usepackage{newtxtext}       %

\usepackage{newtxmath}       

\usepackage{url}

\usepackage{booktabs}

\usepackage{xcolor}

\usepackage{tikz-cd}
\usepackage{tikz}
\usetikzlibrary{arrows,calc,matrix,topaths,positioning,scopes,shapes,decorations, decorations.pathmorphing}
\usepackage{tabularx}
\usepackage{graphicx}
\usepackage{subfigure}
\usepackage{algorithm}
\usepackage{algorithmicx}
\usepackage{algpseudocode}
\usepackage[margin=1in]{geometry}
\usepackage{epsfig} 
\usepackage{comment}
\usepackage{enumitem}

\usepackage{latexsym}

\usepackage{amssymb}

\usepackage{changepage}

\spnewtheorem{theorem}{Theorem}[section]{\bfseries}{}
\spnewtheorem{lemma}[theorem]{Lemma}{\bfseries}{}
\spnewtheorem{corollary}[theorem]{Corollary}{\bfseries}{}
\spnewtheorem{definition}[theorem]{Definition}{\bfseries}{}
\spnewtheorem{remark}[theorem]{Remark}{\bfseries}{}
\spnewtheorem{example}[theorem]{Example}{\bfseries}{}
\spnewtheorem{proposition}[theorem]{Proposition}{\bfseries}{}
\spnewtheorem*{theorem*}{Theorem}{\bfseries}{}
\spnewtheorem*{rule*}{Rule}{\bfseries}{}
\spnewtheorem{notation}[theorem]{Notation}{\bfseries}{}

\usepackage{ntheorem}
\theoremstyle{empty}

\newcommand{\od}{\stackrel{\mbox {\tiny {def}}}{=}}

\def\RR{\mathbb{R}}

\def\RR{\mathbb{R}}

\def\U{\mathcal{U}}
\def\N{\mathcal{N}}

\def\S{{\mathcal{S}}}

\def\P{\mathcal{P}}

\def\RR{\mathbb{R}}

\def\U{\mathcal{U}}
\def\W{\mathcal{W}}
\def\T{\mathcal{T}}

\def\S{{\mathcal{S}}}

\def\P{\mathcal{P}}

\def\max{\mathrm{max}}

\def\supp{\operatorname{supp}}

\def\od{\stackrel{\mathrm{def}}{=}}

\def\supp{\operatorname{supp}}

\def\FP{\operatorname{FP}}

\def\vu{\nu}

\usepackage{color}
\definecolor{blue}{rgb}{0,0,1}
\definecolor{cherry}{rgb}{0.9,.1,.2}

\begin{document}

\title*{Nerve theorems for fixed points of neural networks}
\titlerunning{Nerve theorems for neural networks}

\author{Daniela Egas Santander, Stefania Ebli, Alice Patania, Nicole Sanderson, Felicia Burtscher,\\ Katherine Morrison*, Carina Curto*}
\authorrunning{D.\ Egas Santander, S.\ Ebli, A.\ Patania, N.\ Sanderson, F.\ Burtscher, K.\ Morrison, C.\ Curto}

\institute{
Daniela Egas Santander \at \'Ecole Polytechnique F\'ed\'erale de Lausanne, Lausanne, Switzerland,
\email{daniela.egassantander@epfl.ch}
\and Stefania Ebli \at \'Ecole Polytechnique F\'ed\'erale de Lausanne, Lausanne, Switzerland,
\email{stefania.ebli@epfl.ch}
\and  Alice Patania \at  Indiana University Network Science Institute (IUNI), Bloomington, IN, USA,
\email{apatania@iu.edu}
\and Nicole Sanderson \at Lawrence Berkeley National Lab, Berkeley, CA, USA,
\email{nikki.f.sanderson@gmail.com}
\and Felicia Burtscher\at Technische Universit\"at Berlin, Germany\\
Universit\'e du Luxembourg, Belvaux, Luxembourg (current affiliation)
\email{felicia.burtscher@uni.lu}
\and Katherine Morrison* \at University of Northern Colorado, Greeley, CO, USA,
\email{katherine.morrison@unco.edu}
\and Carina Curto* \at  Pennsylvania State University, University Park, PA, USA,
\email{ccurto@psu.edu }
\and
{*equal contribution}
}

\maketitle

\abstract{ 
Nonlinear network dynamics are notoriously difficult to understand. Here we study a class of recurrent neural networks called combinatorial threshold-linear networks (CTLNs) whose dynamics are determined by the structure of a directed graph. They are a special case of TLNs, a popular framework for modeling neural activity in computational neuroscience. In prior work, CTLNs were found to be surprisingly tractable mathematically. For small networks, the fixed points of the network dynamics can often be completely determined via a series of {\it graph rules} that can be applied directly to the underlying graph. For larger networks, it remains a challenge to understand how the global structure of the network interacts with local properties. In this work, we propose a method of covering graphs of CTLNs with a set of smaller {\it directional graphs} that reflect the local flow of activity. While directional graphs may or may not have a feedforward architecture, their fixed point structure is indicative of feedforward dynamics. The combinatorial structure of the graph cover is captured by the {\it nerve} of the cover. The nerve is a smaller, simpler graph that is more amenable to graphical analysis. We present three nerve theorems that provide strong constraints on the fixed points of the underlying network from the structure of the nerve. We then illustrate the power of these theorems with some examples. Remarkably, we find that the nerve not only constrains the fixed points of CTLNs, but also gives insight into the transient and asymptotic dynamics. This is because the flow of activity in the network tends to follow the edges of the nerve.}
\bigskip
\bigskip
\bigskip
\bigskip
\bigskip
\bigskip
\bigskip
\bigskip

\pagebreak

\section{Introduction}
Combinatorial threshold-linear networks (CTLNs) are a special class of threshold-linear networks (TLNs) whose dynamics are determined by the structure of a directed graph. The firing rates $x_1(t),\ldots,x_n(t)$ of $n$ recurrently-connected neurons evolve in time according to the standard TLN equations:
\begin{equation}\label{eq:dynamics}
\dfrac{dx_i}{dt} = -x_i + \left[\sum_{j=1}^n W_{ij}x_j+\theta_i \right]_+, \quad i = 1,\ldots,n.
\vspace{-.04in}
\end{equation}
These networks derive their name from the nonlinear transfer function, $[\cdot]_+ = \max\{0,\cdot\}$, which is threshold-linear. A given TLN is specified by the choice of a connection strength matrix $W$ and a vector of external inputs $\theta$. TLNs have been widely used in computational neuroscience as a framework for modeling recurrent neural networks, including associative memory networks \cite{AppendixE,Seung-Nature,HahnSeungSlotine, XieHahnSeung,net-encoding, pattern-completion, Fitzgerald2020, Horacio-paper}.

\begin{figure}[!ht]
\vspace{-.1in}
\begin{center}
\includegraphics[width=.9\textwidth]{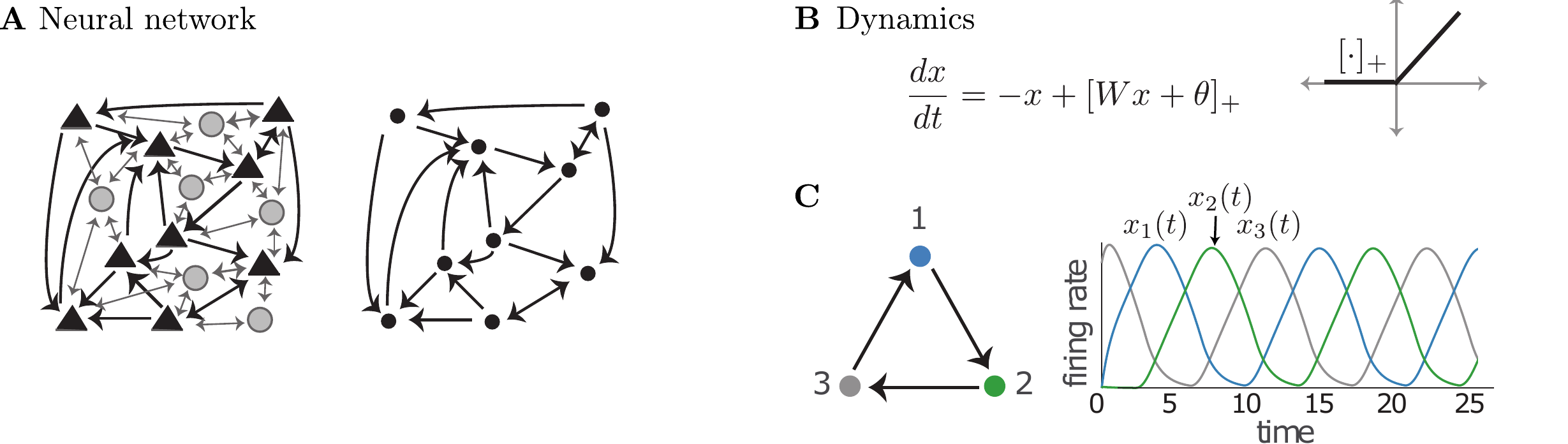}
\caption{(A) A neural network with excitatory pyramidal neurons (triangles) and a background network of inhibitory interneurons (gray circles) that produces a global inhibition. The corresponding graph (right) retains only the excitatory neurons and their connections. (B) TLN dynamics. (C) A graph that is a 3-cycle (left), and a solution for the corresponding CTLN showing that network activity follows the arrows in the graph (right).  Peak activity occurs sequentially in the cyclic order 123. Unless otherwise noted, all simulations have parameters $\varepsilon=0.25, \delta = 0.5,$ and $\theta =1$.}
\label{fig:network-setup}
\end{center}
\vspace{-.25in}
\end{figure}

What makes CTLNs special is that the matrix $W = W(G,\varepsilon,\delta)$ is determined by a simple\footnote{A graph is \emph{simple} if it does not have loops or multiple edges between a pair of vertices.}
 directed graph, as follows:
\begin{equation} \label{eq:binary-synapse}
W_{ij} = \left\{\begin{array}{cc} 0 & \text{ if } i = j, \\ -1 + \varepsilon & \text{ if } j \rightarrow i \text{ in } G,\\ -1 -\delta & \text{ if } j \not\rightarrow i \text{ in } G. \end{array}\right. \quad \quad \quad \quad
\vspace{-.05in}
\end{equation}
Additionally, we fix $\theta_i = \theta > 0$ to be the same for all neurons, and we require the $\varepsilon, \delta$ parameters to satisfy $\delta >0$, and $0 < \varepsilon < \frac{\delta}{\delta+1}$. 
CTLNs were first defined in \cite{CTLN-preprint}, where the $\varepsilon < \frac{\delta}{\delta+1}$ condition was motivated by the desired property that subgraphs consisting of a single directed edge $i \to j$ should not be allowed to support stable fixed points. Note that the upper bound on $\varepsilon$ implies $\varepsilon < 1$, rendering the $W$ matrix effectively inhibitory. We think of the graph edges as excitatory connections in a sea of inhibition (Figure~\ref{fig:network-setup}A). Figure~\ref{fig:network-setup}C shows an example solution for a CTLN whose graph is a $3$-cycle.

TLNs are high-dimensional nonlinear systems whose dynamics are still poorly understood. However, in the special case of CTLNs, there appears to be a strong connection between the attractors of the network and the pattern of stable and unstable fixed points \cite{book-chapter, rule-of-thumb}.\footnote{A fixed point, $x^*,$ of a TLN is a solution that satisfies $dx_i/dt|_{x^*} = 0$ for each $i \in [n]$.} Moreover, these fixed points can often be completely determined by the structure of the underlying graph. In prior work, a series of {\it graph rules} were proven that can be used to determine fixed points of the CTLN by analyzing $G$, irrespective of the choice of parameters $\varepsilon, \delta,$ and $\theta$ \cite{fp-paper, stable-fp-paper}. A key observation is that for a given network, there can be at most one fixed point per support, $\sigma \subseteq [n]$, where the {\it support} of a fixed point is the subset of active neurons (i.e., $\supp{x} = \{i \mid x_i>0\}$).

\pagebreak

\begin{figure}[!ht]
\vspace{-.15in}
\begin{center}
\includegraphics[width=.85\textwidth]{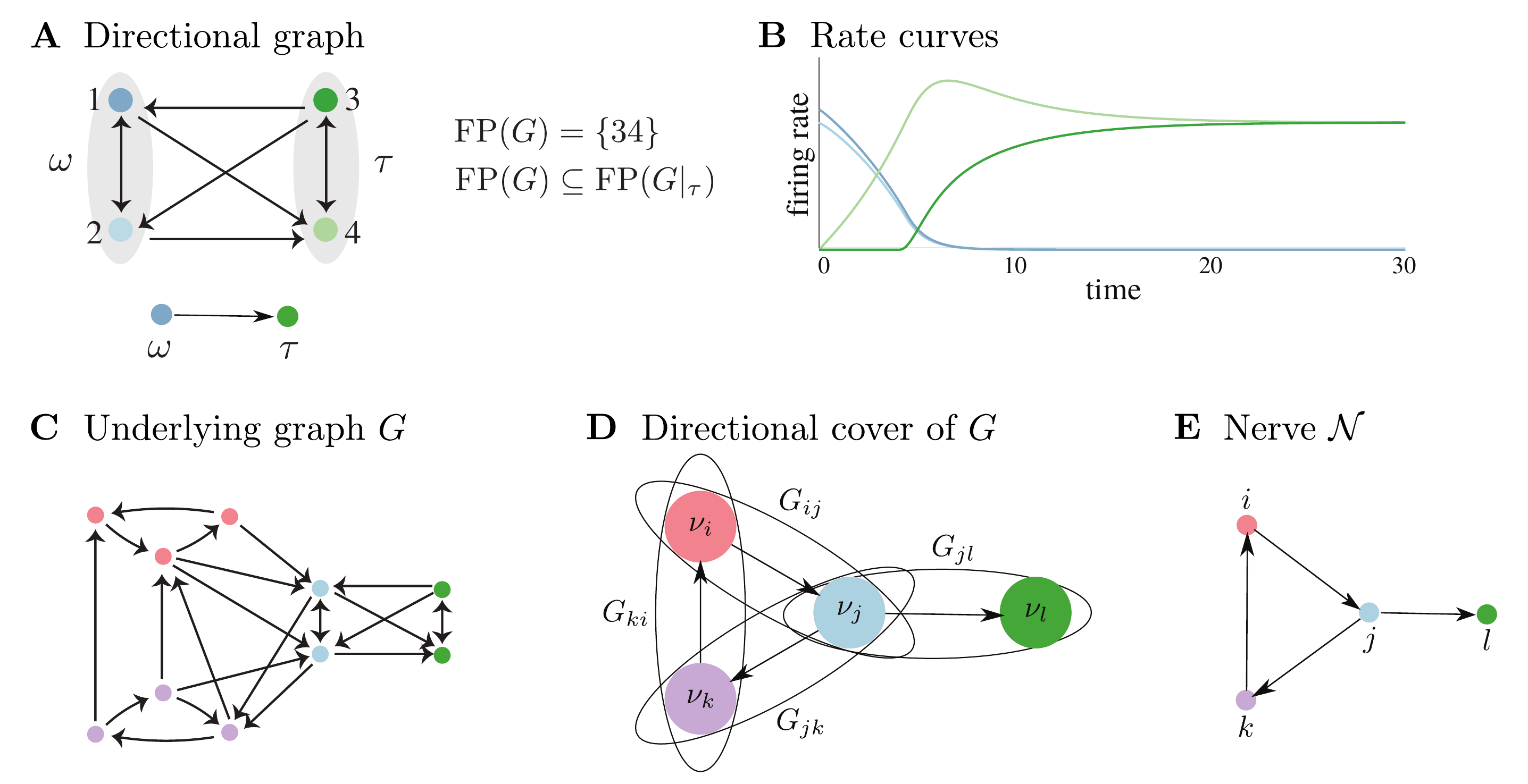}
\caption{Directional graphs and covers. (A) A directional graph, with fixed points supported in the subset of nodes $\tau = \{3,4\}$. (B) A solution of a CTLN with the graph in A. The network was initialized with the activity concentrated on the neurons in $\omega = \{1,2\}$, but the activity flows from $\omega \to \tau$. (C) A graph with a partition of the nodes, each component in a different color. (D) A directional cover of G. Subsets of nodes, $\nu_i, \nu_j, \nu_k,$ and $\nu_l$, correspond to the partition in C. The four induced subgraphs within each oval, of the form $G_{ij} = G|_{\nu_i \cup \nu_j},$ are all directional graphs with direction $\nu_i \to \nu_j,$ as given by the arrows. (E) The nerve associated to the directional cover in D. The corresponding network can be viewed as a dimensional reduction of the one in C.}
\label{fig:new-intro-fig}
\end{center}
\vspace{-.2in}
\end{figure}

For a given choice of parameters, we use the notation 
\[\FP(G) \od \{\sigma \subseteq [n] ~|~  \sigma \text{ is a fixed point support of } W(G,\varepsilon,\delta) \},\]
where $[n] \od \{1, \ldots, n\}$.
For many graphs, we find that the fixed point supports in $\FP(G)$ are confined to a subset of the neurons. In other words, there is a partition $\lbrace\omega,\tau\rbrace$ of the vertices of $G$ such that, for every $\sigma \in \FP(G)$, we have $\sigma \subseteq \tau$ (see Figure~\ref{fig:new-intro-fig}A). In these cases, we observe that solutions of the network activity $x(t)$ tend to converge to a region of the state space where the most active neurons are in $\tau$, and those in $\omega$ are either silent or have very low firing (see Figure~\ref{fig:new-intro-fig}B). In other words, the attractors live where the fixed points live.

This motivates us to define {\it directional graphs}. A directional graph $G$ is a graph with a proper subset of neurons $\tau$ such that $\FP(G) \subseteq \FP(G|_\tau)$, where $G|_\tau$ is the induced subgraph obtained by restricting to the vertices of $\tau$. For example, the graph in Figure~\ref{fig:new-intro-fig}A is directional with a single fixed point supported in $\tau = \{3,4\}$. We also require an additional technical condition that allows us to prove that certain natural compositions, like chaining directional graphs together, produce a new directional graph 
(see Definition \ref{def:directional} for the full definition). In simulations, such as the one in Figure~\ref{fig:new-intro-fig}B, we have seen that directional graphs display feedforward dynamics, even if their architecture does not follow a feedforward structure. Activity that is initially concentrated on $\omega$ flows towards $\tau$, giving the dynamics an $\omega\to\tau$ directionality.  
Thus, from a bird's eye view, directional graphs behave like a single directed edge, where the activity flows from the source to the sink. These observations prompted us to ask the following question:
if we cover a graph $G$ with a collection of directional graphs, what can we say about $\FP(G)$ from the combinatorial structure of the cover?

In this paper we develop tools to answer this question, inspired by the construction of the nerve of a cover of a topological space. We define a {\it directional cover} of $G$ as a set of directional subgraphs that cover $G$ and  have well-behaved intersections.\footnote{
This is analogous to the definition of a ``good" cover of a topological space, which also requires well-behaved intersections.
Nerves of good covers reflect the topology of the underlying space \cite{borsuk1948imbedding,leray1945forme}.}
Effectively, such a cover is entirely determined by a partition of the vertices of $G$, denoted $\{\nu_i\}$, that satisfies special properties. 
(See Definition \ref{def:directional_cover} for a precise definition.)
We define the {\it nerve} of a directional cover as a new graph $\N$ that has a directed edge for each directional graph in the cover, and a vertex for each component $\nu_i$ of the partition.
Figure~\ref{fig:new-intro-fig}C,D depicts a graph with a partition of the nodes (indicated by the colors), and its corresponding directional cover.
The edges of the nerve reflect the local dynamics of $G$, and the 
nerve itself encodes the combinatorics of the intersection pattern of the cover:
the directional graphs overlap precisely at vertices of the nerve where their edges meet.
The partition of the vertices of $G$ induces a canonical quotient map, 
$\pi: V_G \to V_\N:=\{\nu_i\},$ that simply identifies all the vertices in each component $\nu_i$. 
Figure~\ref{fig:new-intro-fig}E is the nerve of the directional cover in D, and the quotient map $\pi$ sends each node in C to the corresponding node with the same color in E -within the nerve $\N$, we often label the node $\nu_i$ simply as $i$.

\begin{figure}[!ht]
\vspace{-.05in}
\begin{center}
\includegraphics[width=.9\textwidth]{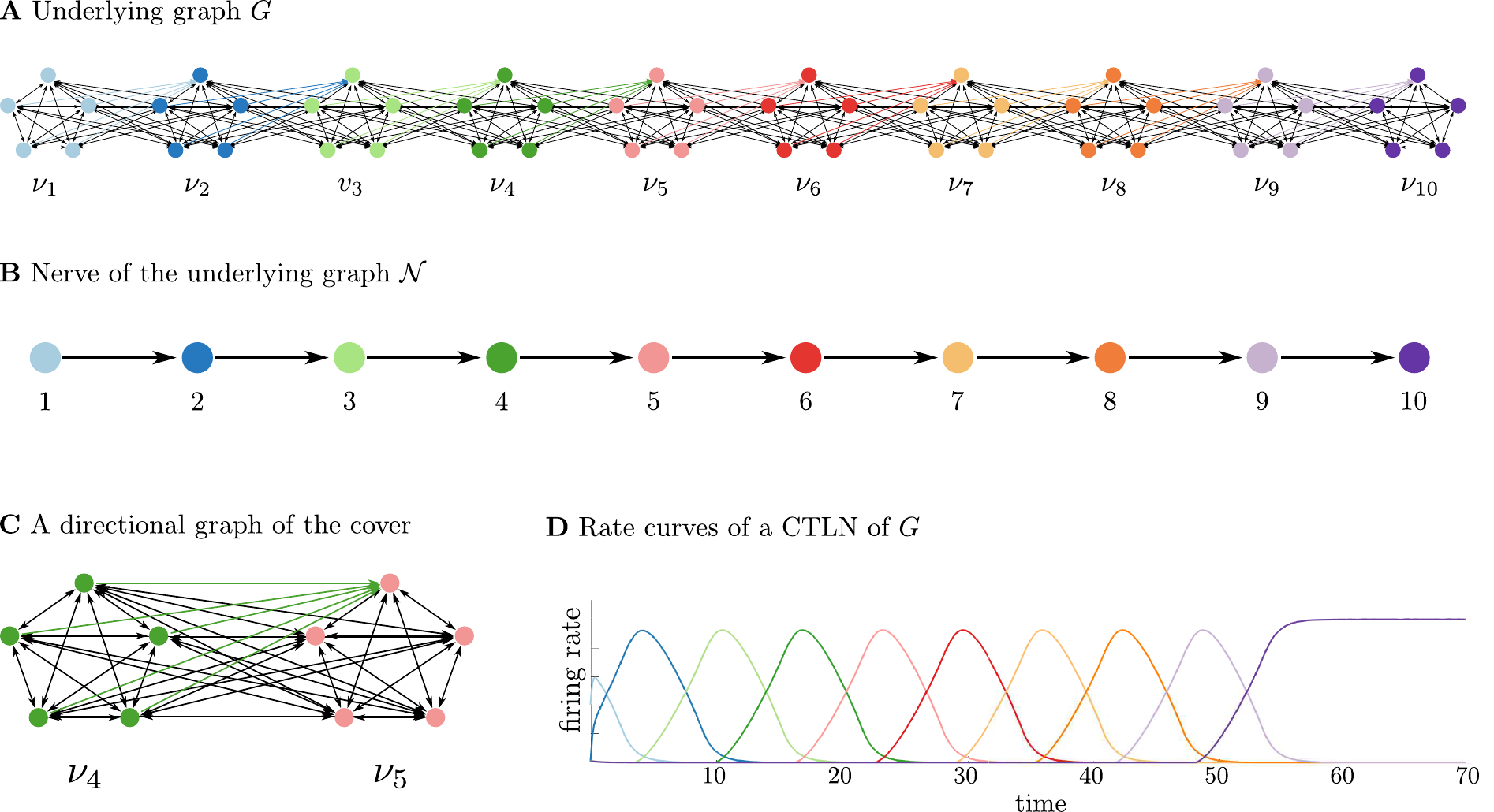}
\caption{{\bf Example graph with a directional cover, its nerve, and network activity that flows along the nerve.} (A) A chain of ten 5-cliques where the edges between adjacent cliques all follow the pattern shown in panel C.  (B) The nerve of the graph in A induced by the partition of the vertices as $\nu_1, \ldots, \nu_{10}$.  (C) The graph $G$ restricted to a pair of adjacent cliques.  All edges in black are bidirectional, while those in green are unidirectional from the green clique to the pink clique.  This restricted graph is directional, and all the graphs in the directional cover of $G$ have this form.   (D) A solution to the CTLN defined by $G$ (with $\varepsilon=0.25$ and $\delta = 0.5$), where the activity is initialized on the nodes in the first clique $\nu_1$.  The transient dynamics slowly activate each clique in sequence, following the path of the nerve, until the solution converges to the stable fixed point supported on the nodes in $\nu_{10}$.
}
\label{fig:clique-chain}
\end{center}
\vspace{-.2in}
\end{figure}

As an illustration of directional covers and nerves, consider the graph in Figure~\ref{fig:clique-chain}A.  This graph is a chain of ten 5-cliques where the edges between adjacent cliques all follow the pattern shown in panel C: there are edges forward from every node in the first clique to every node in the second clique; every node in the second clique (except for the top node) sends edges back to every node in the first clique.  Most edges are thus bidirectional arrows (in black), while the edges that only go forward from clique i to clique i+1 are in color.  
The induced subgraphs $G|_{\nu_i \cup \nu_{i+1}}$ are all directional with direction $\nu_i \to \nu_{i+1}$, despite all the back edges from right to left. This means that $\FP(G|_{\nu_i \cup \nu_{i+1}}) \subseteq \FP(G|_{\nu_{i+1}})$, and we expect the activity of the neurons to flow from $\nu_i $ to $\nu_{i+1}$ (left to right).   Figure~\ref{fig:clique-chain}B depicts the nerve of $G$. Figure~\ref{fig:clique-chain}D shows the solution to a CTLN defined by $G$, where we have initialized all the activity on the nodes in the first clique $\nu_1$.  We see that the activity eventually converges to the final component, shown in purple, where the fixed points of the network are concentrated. The transient dynamics, however, are rather slow, with each clique activated in a sequence that follows the path-like structure of the nerve. Note that this network behaves similarly to a synfire chain \cite{synfire-chain1, synfire-chain2, synfire-chain3}, despite numerous backward edges between components that completely destroy the feedforward architecture (Figure~\ref{fig:clique-chain}C). The sequential dynamics are maintained because these backward edges do not disrupt the directionality of the graphs in the cover.

The main goal of this paper is to prove nerve theorems for CTLNs. Broadly speaking, such a {\it nerve theorem} is a result that gives information about the dynamics of a network from properties of the nerve. Specifically, we are interested in results that allow us to constrain the fixed points of $G$ by analyzing structural properties of $\N$. 
Ideally, we would like to prove the following kind of result: If $G$ has a directional cover with nerve $\N$, then 
\begin{equation}\label{eq:wishful}
\sigma \in \FP(G) \; \Rightarrow \; \pi(\sigma) \in \FP(\N),
\end{equation}
where $\pi$ is the canonical quotient map from $V_G$ to $V_{\N}$.
This can be quite powerful in cases where $\N$ is a much smaller and simpler graph. 
Unfortunately, the statement~\eqref{eq:wishful} is not in general true. However, we do find that this holds whenever the nerve $\N$ is a directed acyclic graph (DAG) or $\N$ is a cycle (see Theorems~\ref{thm:DAG} and~\ref{thm:cycle-nerve}).  More generally, whenever $\N$ admits a DAG decomposition (see Definition~\ref{def:DAG-decomp}), Theorem \ref{thm:DAG-decomp} gives a result similar in spirit to~\eqref{eq:wishful} and allows us to greatly constrain $\FP(G)$.

Our nerve theorems can be used to simplify a complex network by finding a nontrivial directional cover and studying its nerve. Finding such covers is an art, however, and we do not yet have a systematic way of doing it. On the other hand, nerve theorems can also be used to engineer complex networks with prescribed dynamic properties. This is how we constructed the example in Figure~\ref{fig:clique-chain}. We explore both kinds of applications in the last section of the paper.

The organization of this paper is as follows. In Section~\ref{sec:background}, we review some graph theory terminology and basic background and notation for CTLNs. We also introduce the DAG decomposition of a graph.  In Section~\ref{sec:dir-graphs}, we define directional graphs, prove that certain graph structures are always directional, and provide some other families of examples. In Section~\ref{sec:dir-covers}, we introduce directional covers and their associated nerves. Here we also state and prove our main results, Theorems~\ref{thm:DAG-decomp},~\ref{thm:DAG} and~\ref{thm:cycle-nerve}. Finally, in Section~\ref{sec:applications}, we illustrate the power of our theorems with some applications.

\section{Preliminaries}\label{sec:background}
In this section we review some useful terminology from graph theory and summarize essential background and prior results about fixed points of CTLNs. We also introduce the DAG decomposition of a graph, a notion that will appear in our main nerve theorems.

\subsection{Graph theory terminology}

\begin{definition}
A \textit{directed graph} $G$ can be described as a tuple $G=(V_G,E_G)$, where $V_G$ is a finite set called the set of \textit{vertices} and 
$E_G\subseteq V_G\times V_G$ is the the set of \textit{(directed) edges}, where $(i,j) \in E_G$ means there is a directed edge $i \to j$ from $i$ to $j$ in $G$.  If $(i,j) \notin E_G$, we write $i \not\to j$. A directed graph is \textit{simple} if it has no self-loops, so that $(i,i) \notin E_G$ for all $i \in V_G$.  A directed graph is \textit{oriented} if it has no bidirectional edges.
\end{definition}

In this paper, we restrict ourselves to simple directed graphs.   Unless otherwise noted, we will use the word {\it graph} to refer to simple directed graphs. 

\begin{notation}
Let $G$ be a graph with vertex set $V_G$ and edge set $E_G$.
For any subset of vertices $\sigma \subseteq V_G$, denote by $G|_{\sigma}$ the \textit{induced subgraph} obtained by restricting to the vertices $\sigma$.  More precisely, $G|_\sigma = (\sigma,E|_{\sigma})$ where $E|_{\sigma} = \{(i,j) \in E_G \mid i,j \in \sigma\}$.

Let $\sigma_1,\sigma_2 \subseteq V_G$ be two subsets of the vertices of $G$.  We denote by $E_G(\sigma_1,\sigma_2)\subseteq E_G$ the set of directed edges \textit{from} vertices in $\sigma_1$  \textit{to} vertices in $\sigma_2$ in $G$.
\end{notation}

Next, we define some basic notions relevant to graphs. 

\begin{definition}
Let $G$ be a graph and $v\in V_G$ be a vertex in $G$. The \textit{in-degree} of $v$ is the number of incoming edges to $v$. The \textit{out-degree} of $v$ is the number of outgoing edges from $v$.
We say $v$ is a \textit{source} if $v$ has no incoming edges, and we say $v$ is a \textit{proper source} if it is a source that has at least one outgoing edge.
We say $v$ is a \textit{sink} if $v$ has no outgoing edges.
Note that a source that is not a proper source is an \textit{isolated vertex}, and thus it is also a sink.  
\end{definition}

\begin{definition}
We say that a graph $G$ has \textit{uniform in-degree} if every vertex $v \in V_G$ has the same in-degree $d$. Note that an {\it independent set} is a graph with uniform in-degree $d = 0$. A {\it $k$-clique} is an all-to-all bidirectionally connected graph with uniform in-degree $d = k-1$. And an {\it $n$-cycle} is a graph with $n$ edges, $1 \to 2 \to \cdots \to n \to 1$, which has uniform in-degree $d = 1$.
\end{definition}

\subsection{Background on fixed points of CTLNs}

In this subsection we recall the results from \cite{fp-paper} that are relevant for this work and include simple proofs to some of these to provide intuition to the reader.

A {\it fixed point} of a CTLN is simply a fixed point of the network equations~\eqref{eq:dynamics}. In other words, it is a vector $x^* \in \RR_{\geq 0}^n$ such that $\dfrac{dx_i}{dt}|_{x = x^*} = 0$ for all $i \in [n]$. As explained in \cite{fp-paper}, fixed points of CTLNs can be labelled by their {\it supports} (i.e. the subset of active neurons),
and for a given $G$ the set of all fixed point supports is denoted $\FP(G)$.

\begin{lemma}[\cite{fp-paper}]\label{lemma:uniform-in-degree}
Let $G$ be a graph on $n$ vertices, and suppose $G$ has uniform in-degree. Then $G$ has a full-support fixed point, $\sigma = [n] \in \FP(G)$.
\end{lemma}

In particular, this lemma says that cliques, cycles, and independent sets all have a full-support fixed point. In fact, this fixed point is symmetric, with $x_i^* = x_j^*$ for all $i,j \in [n]$. This is true even for uniform in-degree graphs that are not symmetric.

More generally, fixed points can have very different values across neurons. However, there is some level of ``graphical balance'' that is required of $G|_\sigma$ for any fixed point support $\sigma$. For example, if $\sigma$ contains a pair of neurons $j,k$ that have the property that all neurons mapping to $j$ are also mapping to $k$, and $j \to k$ but $k \not\to j$, then $\sigma$ cannot be a fixed point support. This is because $k$ is receiving strictly more inputs than $j$, and this imbalance rules out their ability to coexist in the same fixed point support. To see this more rigorously, we have the following lemma.

\begin{lemma}\label{lemma:domination}
Let $G$ be a CTLN and $\sigma \subseteq V_G$. Suppose there exist vertices $j,k \in \sigma$ such that for each $i \in \sigma\setminus\{j,k\},$ if $i \to j$ then $i \to k$. Furthermore, suppose $j \to k$ but $k \not\to j$. Then $\sigma \notin \FP(G)$.
\end{lemma}

\begin{proof}
To obtain a contradiction, assume $\sigma \in \FP(G)$. The corresponding fixed point $x$ satisfies $x_i > 0$ for all $i \in \sigma$, and $dx_i/dt = 0$. In particular, setting $dx_j/dt = 0$ and $dx_k/dt = 0$ (and recalling $W_{jj} = W_{kk} = 0$) we obtain:
\begin{eqnarray*}
x_j &=& \sum_{i \in \sigma\setminus\{j,k\}} W_{ji} x_i + W_{jk}x_k + \theta,\\
x_k &=& \sum_{i \in \sigma\setminus\{j,k\}} W_{ki} x_i + W_{kj}x_j + \theta.
\end{eqnarray*}
Now observe that for each $i \in \sigma\setminus\{j,k\}$, the fact that $i \to j$ implies $i \to k$ tells us that $W_{ji} \leq W_{ki}$, (see Equation \eqref{eq:binary-synapse}).
This means the summation term in the $x_j$ equation above is less than or equal to the analogous term in the $x_k$ equation. Using this fact, we see that $x_j - W_{jk}x_k \leq x_k - W_{kj}x_j$, which can be rearranged as,
$$(1+W_{kj})x_j \leq (1+W_{jk})x_k.$$
Now recall that $j \to k$ but $k \not\to j$, so $W_{kj} = -1+\varepsilon$ and $W_{jk} = -1-\delta$. The above inequality thus says that $\varepsilon x_j \leq -\delta x_k$. But this is a contradiction, because $\varepsilon x_j>0$ and $-\delta x_k < 0$.
And so no fixed point supported on $\sigma$ can exist.
\end{proof}

The conditions on $j,k \in \sigma$ used in the above lemma is an example of so-called {\it graphical domination}. This notion was first defined in \cite{fp-paper}, and provides a useful tool for ruling in and ruling out fixed points of CTLNs purely based on the graph structure, and independently of the $\varepsilon, \delta$ and $\theta$ parameters.  

\begin{definition}[graphical domination]\label{def:domination}
Let $G$ be a graph, $\sigma \subseteq V_G$ a subset of the vertices, and $j, k \in V_G$ such that $\{j,k\}\cap \sigma\neq\emptyset$. 
We say that $k$ \textit{graphically dominates} $j$ with respect to $\sigma$,  and write $k >_{\sigma} j$, if the following three conditions hold:
\begin{enumerate}
\item For all $i \in \sigma\setminus \{j,k\}$ if $i\rightarrow j$, then $i \rightarrow k$.
\item If $j \in \sigma$, then $j\rightarrow k$.
\item If $k \in \sigma$, then $k\nrightarrow j$.
\end{enumerate}
\end{definition}

This definition of domination covers more cases than what we saw in Lemma~\ref{lemma:domination}. This greater generality is reflected in the main theorem about domination, which appeared as Theorem 4 in \cite{fp-paper}.  We cite a special case of this theorem below.

\begin{theorem}[graphical domination \cite{fp-paper}]\label{thm:graphdomination}
Let $\sigma\subseteq V_G$ be a subset of the vertices of a graph $G$.  If there is a $j\in\sigma$ and a $k\in V_G$ such that $k >_{\sigma} j$ ($k$ graphically dominates $j$ with respect to $\sigma$), then $\sigma \notin \FP(G)$.
\end{theorem}

\begin{figure}[hbtp]
  \centering
  \includegraphics[width=90mm]{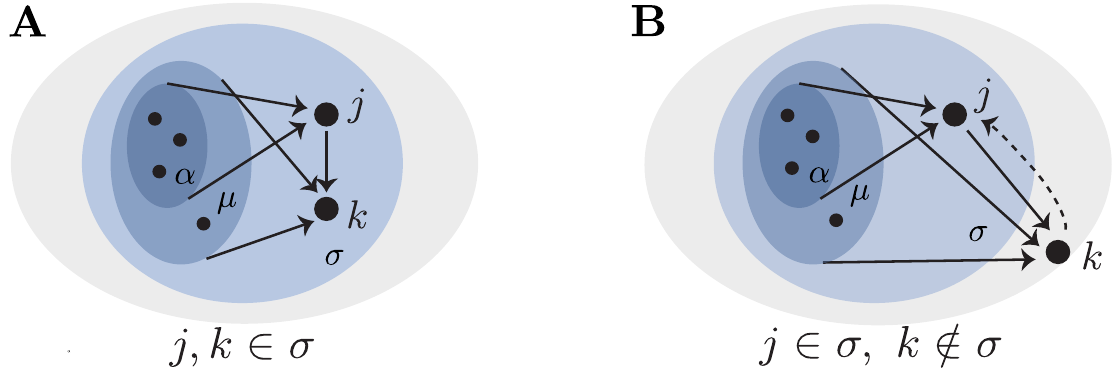}
\caption{{\bf The two cases of graphical domination in Theorem~\ref{thm:graphdomination}.} 
In each panel, $k$ graphically dominates $j$ with respect to $\sigma$ (the outermost shaded region). The inner shaded regions illustrate the subsets of nodes that send edges to $j$ and $k$. Note that the vertices sending edges to $j$ are a subset of those sending edges to $k$, but this containment need not be strict. The dashed arrow indicates an optional edge between $j$ and $k$.}
  \label{fig:graph_domination}
\end{figure}

We will furthermore use the following useful equivalence, which states that $\sigma$ can only be a fixed point support if $\sigma \in \FP(G|_\sigma)$ and the fixed point survives the addition of each individual $k \notin \sigma.$

\begin{lemma}[{\cite[Corollary 2]{fp-paper}}]\label{lemma:cor2}
Consider a CTLN determined by a graph $G$ on a set of neurons $V_G$, and let $\sigma \subseteq V_G$.  Then
\[\sigma\in\FP(G) \ \ 
\Leftrightarrow \ \ 
\sigma \in \FP (G|_{\sigma \cup \{k\}}) \text{ for all } k \in V_G.\]
In particular,  $\sigma \in \FP(G) \Rightarrow \sigma \in \FP(G|_\sigma)$.  Moreover, $\sigma \in \FP(G) \Rightarrow \sigma \in \FP(G|_\tau)$ for any $\tau$ with $\sigma \subseteq \tau$.  
\end{lemma}

One simple case where graphical domination can be used to rule out a fixed point support is whenever a graph contains a proper source. This is Rule 6 in \cite{fp-paper}.

\begin{lemma}[sources \cite{fp-paper}]\label{lemma:sources}
Let $G$ be a graph and $\sigma \subseteq V_G$.  If there exists a $j \in \sigma$ such that $j$ is a proper source in $G|_{\sigma}$ or $j$ is a proper source in $G|_{\sigma \cup \{\ell\}}$ for some $\ell \in V_G$, then $\sigma \notin \FP(G)$.
\end{lemma}

\begin{proof}
If $j$ is a proper source in $G|_{\sigma}$, then there exists $k \in \sigma$ such that $j \to k$. Since $j$ has no other inputs in $\sigma$, clearly $k >_\sigma j.$ If $j$ is not a proper source in $G|_{\sigma}$  but is a proper source in $G|_{\sigma \cup \ell}$, then $j \to \ell$, and hence $\ell >_\sigma j$. In either case, by Theorem~\ref{thm:graphdomination} we have that $\sigma \notin \FP(G)$.  
\end{proof}

The lemma above allows us to rule out fixed points of cycles that are not full support. 

\begin{lemma}[cycles]\label{lemma:FP-cycle}
If $G$ is a cycle on $n$ vertices, then $G$ has a unique fixed point, which has full support.  In other words,
$$\FP(G) = \{ [n]\}.$$
\end{lemma}
\begin{proof}
First observe that $[n] \in \FP(G)$ by Lemma~\ref{lemma:uniform-in-degree} because a cycle has uniform in-degree 1.  To see that this is the only fixed point support of $G$, consider any proper subset $\sigma \subsetneq V_G$.  Since $G$ is a cycle, $G|_\sigma$ either contains a path or is an independent set.  If it contains a path, then the source of that path is a proper source in $G|_\sigma$.  If it is an independent set, then for any $i \in \sigma$, we have $i \to \ell$ in $G$ for $\ell = i+1$.  Then $i$ is a proper source in $G|_{\sigma \cup \ell}$.  Thus by Lemma~\ref{lemma:sources}, $\sigma \notin \FP(G)$.  Thus, $\FP(G) = \{ [n]\}.$
\end{proof}

Another simple case when graphical domination can be used to rule out fixed points is whenever $\sigma$ has a target in $G$.  For $k \in V_G$, we say that $k$ is a \emph{target} of $\sigma$ if $i \to k$ for every $i \in \sigma\setminus\{k\}$.

\begin{lemma}[targets \cite{fp-paper}]\label{lemma:targets}
Let $G$ be a graph and $\sigma \subseteq V_G$. Suppose $k \in V_G$ is a target of $\sigma$.
\begin{enumerate}
\item If $k \in V_G \setminus \sigma$, then $\sigma \notin \FP(G)$.
\item If $k \in \sigma$ and there exists a $j \in \sigma$ such that $k \not\to j$, then $\sigma \notin \FP(G)$.
\end{enumerate}
\end{lemma}
\begin{proof}
In case 1, it is straightforward to see that $k>_\sigma j$ for any $j \in \sigma$.  In case 2, we see that for the particular $j$ such that $k \not\to j$, we have $k >_\sigma j$.  In either case, by Theorem~\ref{thm:graphdomination} we have that $\sigma \notin \FP(G)$. 
\end{proof}

The target lemma allows us to rule out fixed points of cliques that are not full support.

\begin{lemma}[cliques]\label{lemma:FP-clique}
If $G$ is a clique on $n$ vertices, then $G$ has a unique fixed point, which has full support.  In other words,
$$\FP(G) = \{ [n]\}.$$
\end{lemma}
\begin{proof}
First observe that $[n] \in \FP(G)$ by Lemma~\ref{lemma:uniform-in-degree} because a clique has uniform in-degree $n-1$.  To see that this is the only fixed point support of $G$, consider any proper subset $\sigma \subsetneq V_G$ and let $k \in V_G \setminus \sigma$.  Then $k$ is a target of $\sigma$, and so by Lemma~\ref{lemma:targets}, $\sigma \notin \FP(G)$.  Thus, $\FP(G) = \{ [n]\}.$
\end{proof}

Finally, using a more general form of domination defined in \cite{fp-paper}, we obtain the following {\it survival rule} telling us precisely when a uniform in-degree fixed point survives as a fixed point of a larger network (Theorem 5 of \cite{fp-paper}):

\begin{theorem}[uniform in-degree \cite{fp-paper}] \label{thm:uniform-in-degree}
Let $G$ be a graph and $\sigma \subseteq V_G$ such that $G|_\sigma$ has uniform in-degree $d$. For $k \in V_G \setminus \sigma$, let $d_k \od |\{i \in \sigma \mid i \to k\}|$ be the number of edges $k$ receives from $\sigma$.  Then 
$$ \sigma \in \FP(G) \ \ \Leftrightarrow \ \  d_k \leq d \text{ for every } k \in V_G \setminus \sigma.$$
\end{theorem}

Other than this theorem we will not use the more general form of domination.  Therefore, in the remainder of this work when we say domination we mean graphical domination.
\subsection{The DAG decomposition}

Two of our main results are nerve lemmas involving {\it directed acyclic graphs} (DAGs). Recall that a DAG is a graph that has no directed cycles. There is a well known characterization of DAGs in terms of a \textit{topological ordering} of their vertices.  In particular, $G$ is a DAG if and only if there exists an ordering of the vertices such that edges in $G$ only go from lower numbered to higher numbered vertices. In other words, if $i \to j$ then $i < j$; equivalently if $i > j$ then $i \not\to j$.

\begin{lemma}[DAGs]\label{lemma:FP-DAG}
Let $G$ be a DAG and let $\tau =\{\text{sinks of } G\}$.  Then the fixed point supports of $G$ are all the nonempty subsets of $\tau$, i.e.
$$\FP(G) = \P(\tau)\setminus\{\emptyset\},$$
where $\P(\tau)$ denotes the power set of $\tau$.  
\end{lemma}
\begin{proof}
First to see that $\P(\tau)\setminus\{\emptyset\} \subseteq \FP(G)$, notice that any non-empty subset of $\tau$ is an independent set of sinks.  An independent set has uniform in-degree 0, and thus by Theorem~\ref{thm:uniform-in-degree}, an independent set produces a fixed point when it has no outgoing edges.  Since all the nodes in $\tau$ are sinks, every subset of $\tau$ has no outgoing edges, and so every subset produces a fixed point support in $\FP(G)$.  

Next, to see that no other sets can produce fixed points of $G$, consider $\sigma \subseteq V_G$ such that $\sigma \not\subseteq \tau$.  Let $j$ be the lowest number vertex in $\sigma \setminus \tau$ according to some topological ordering of $G$.  Then $j$ has no incoming edges from other nodes in $\sigma$ since edges in a DAG can only go from lower numbered vertices to higher number vertices.  Moreover, there exists some $\ell \in V_G$ such that $j \to \ell$ since otherwise $j$ would be a sink, but $j \notin \tau$ by design, which contains all the sinks of $G$.  Thus $j$ is a proper source in $G|_{\sigma \cup \ell}$, and so $\sigma \notin \FP(G)$ by Lemma~\ref{lemma:sources}.
\end{proof}

Many graphs that are not DAGs nevertheless have a DAG-like structure on a subgraph. This will also be a useful concept for our nerve theorems.

\begin{definition}[DAG decomposition]\label{def:DAG-decomp}
Let $G$ be a graph.  For $\omega, \tau \subseteq V_G$, we say that $(\omega, \tau)$ is a \textit{DAG decomposition} of $G$ if $\omega \dot\cup  \tau$ is a partition of the vertices $V_G$ such that: 
\begin{enumerate}
\item $G|_{\omega}$ is a DAG,
\item $G|_\tau$ contains all sinks of $G$, 
\item there are no edges from $\tau$ back to $\omega$, i.e., $E_G(\tau,\omega)=\emptyset$.
\end{enumerate}
We say a DAG decomposition is \textit{non-trivial} if $\omega\neq\emptyset$. 
We say a DAG decomposition is \textit{maximal} if $\omega$ is as large as possible.  More precisely, $(\omega,\tau)$  is a maximal DAG decomposition if there is no other DAG decomposition $(\omega', \tau')$ with $\omega \subsetneq \omega'$.
\end{definition}

\begin{figure}[!h]
\vspace{-.05in}
\begin{center}
\includegraphics[width=.8\textwidth]{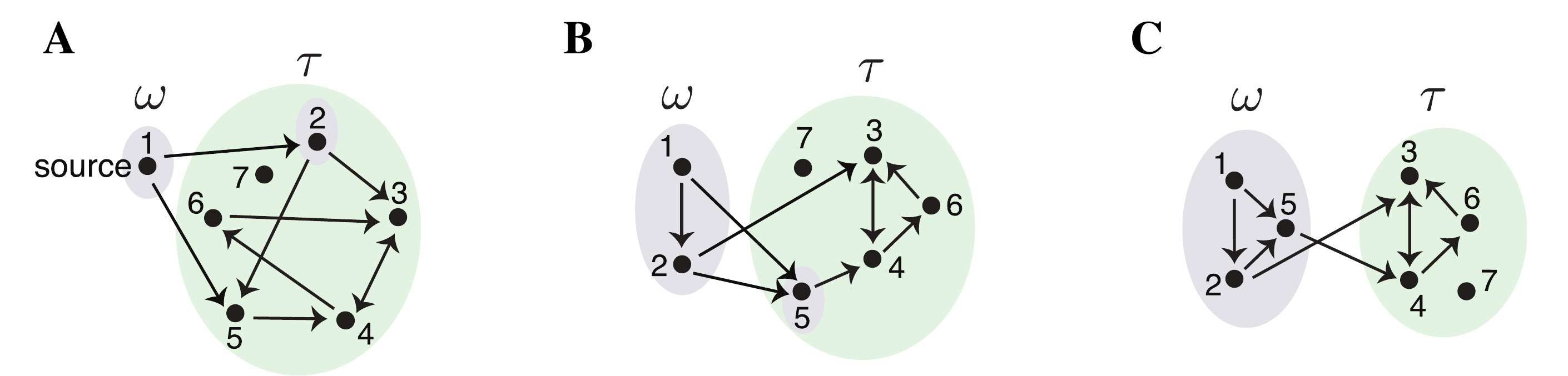}
\vspace{.1in}
\caption{Iterative construction of DAG decompositions. (A) A DAG decomposition of a graph where $\omega$ contains only a single source.  The gray highlighted node $2$ is a proper source in $G|_\tau$, but not a source in the full graph.  (B) A second DAG decomposition is obtained by moving node $2$ to $\omega$. Now node $5$ has become a proper source in the new $G|_\tau$. (C) A third DAG decomposition is obtained by moving $5$ to $\omega$. In this decomposition, $G|_\tau$ has no proper sources. Notice that node 7 is a source, but because it has no outgoing edges, it is not a proper source so will not be moved to $\omega$.  In fact, node 7 is a sink, and thus is required to be in $\tau$ by condition 2 of DAG decompositions.  We have thus arrived at the unique DAG decomposition with minimal $\tau$ and maximal DAG $\omega$.}
\label{fig:DAG-decomp}
\end{center}
\vspace{-.2in}
\end{figure}

Every graph $G$ that has at least one proper source $j$ has a DAG decomposition with $\omega = \{j\}$ and $\tau = V_G \setminus  \{j\}$.  But DAG decompositions are most valuable when $\tau$ is as small as possible.  To minimize the size of $\tau$, we'd like to ``grow" $\omega$ as much as possible, as in a maximal DAG decomposition.  It turns out that there is straight-forward procedure for generating a maximal DAG decomposition of a graph, and moreover, the maximal DAG decomposition is in fact unique.  Specifically, one can iteratively refine a DAG decomposition by moving any nodes that are proper sources in $G|_\tau$ to $\omega$ (see Figure~\ref{fig:DAG-decomp}).  This process will maintain the property that $G|_\omega$ is a DAG (each node that is moved to $\omega$ will be at the end of the ``topological ordering" of the DAG) while also guaranteeing that there are no edges from nodes in $\tau$ back to nodes in $\omega$.   Finally, the process terminates when there are no nodes in $\tau$ that are proper sources in $G|_\tau$.  It turns out that the $\tau$ satisfying $G|_\tau$ has no proper sources is both \textit{minimal}, in the sense that $|\tau|$ is smallest and $\tau \subseteq \tau'$ for any other DAG decomposition $(\omega', \tau')$, and \textit{unique}.  As a result, this process yields the unique maximal DAG decomposition.
Note in particular that in any maximal DAG decomposition $(\omega,\tau)$, $\tau$ cannot have proper sources, because if it did one could move such a vertex to $\omega$, contradicting maximality.

\begin{lemma}\label{lemma:unique-DAG-decomp}
Suppose that $G$ contains a proper source.  Then the DAG decomposition $(\omega, \tau)$ of $G$ satisfying $G|_\tau$ has no proper sources is a maximal DAG decomposition.  In particular, $G$ has a unique maximal DAG decomposition.
\end{lemma}
\begin{proof}
Let $(\omega, \tau)$ be a DAG decomposition of $G$ satisfying $G|_\tau$ has no proper sources, and let $(\omega', \tau')$ be any other DAG decomposition of $G$.  Suppose $\omega' \not\subseteq \omega$.  Then since each DAG decomposition is a partition of the vertices, this condition on $\omega$ implies that $\tau \not\subseteq \tau'$.  Then there exists a node $i_0 \in \tau \setminus  \tau'$.  Since $i_0 \in \tau$ and $G|_\tau$ has no proper sources, there exists some $i_1 \in \tau$ such that $i_1 \to i_0$.  Since $i_1$ also is not a proper source in $G|_\tau$, there exists some $i_2 \in \tau$ such that $i_2 \to i_1 \to i_0$.  Again $i_2$ is not a proper source, and so there exists $i_3 \in \tau$ such that $i_3 \to i_2 \to i_1 \to i_0$.  

Note that in the other DAG decomposition $(\omega', \tau')$, since $i_0 \notin \tau'$, we must have $i_0 \in \omega'$.  Moreover, by the definition of DAG decomposition, there are no edges from nodes in $\tau'$ to nodes in $\omega'$, and so all nodes in the path $i_3 \to i_2 \to i_1 \to i_0$ must also be in $\omega'$.  
We can continue to trace the path backwards in this way through $G|_\tau$ for arbitrarily many steps since it has no proper sources, but since $\tau$ is finite, at some point some node must appear twice in this path.
Thus this sequence of nodes must contain a bidirectional edges and/or a directed cycle.  But all the nodes in this sequence must be in $\omega'$, and by definition, $G|_{\omega'}$ must be a DAG, thus it cannot contain any bidirectional edges or directed cycles.  Thus we have a contradiction, and so $\tau \subseteq \tau'$, and thus $\omega' \subseteq \omega$.  

Since any DAG decomposition $(\omega, \tau)$ of $G$ satisfying $G|_\tau$ has no proper sources must have maximal $\omega$, it follows that there must be a unique decomposition satisfying this property.  Finally, since any maximal DAG decomposition must satisfy this property, it follows there is a unique one and it is this one.
\end{proof}
\section{Directional graphs}\label{sec:dir-graphs}
In this section, we focus on a special class of graphs known as \textit{directional graphs}, first defined in \cite{graph-rules-paper}.  The motivating heuristic behind \textit{directional graphs} is that they are graphs whose vertices can be partitioned into two sets $\omega$ and $\tau$ such that when the neural activity is initialized on nodes in $\omega$, it flows to the nodes in $\tau$.  
In simulations, we have seen that this flow of activity occurs whenever the fixed points of $G$ are confined to live in $\tau$, so that $\FP(G) \subseteq \FP(G|_\tau)$.
In order to guarantee nice properties when we union together directional graphs, we require something slightly stronger in our definition of directional graphs, namely that the collapse of the fixed points onto the subnetwork $G|_\tau$ be the result of graphical domination.

\begin{definition}[directional graph]\label{def:directional}
We say that a graph $G$ is \textit{directional,} with direction $\omega \to \tau$, if $\omega \dot\cup \tau = V_G$ is a nontrivial partition of the vertices ($\omega, \tau \neq \emptyset$, $\omega \cap \tau=\emptyset$) such that $\FP(G) \subseteq \FP(G|_\tau)$ by way of graphical domination.  Specifically, we require the following property: for every $\sigma \not\subseteq \tau$, there exists some $j \in \sigma \cap \omega$ and $k \in V_G$ such that $k$ graphically dominates $j$ with respect to $\sigma$, i.e.\ $k >_\sigma j$.  
When this is the case we say $\sigma$ \textit{dies by (graphical) domination.}
\end{definition}

As mentioned above, we predict that directional graphs will have {\it feedforward dynamics}, so that activity that is initially concentrated on $G|_\omega$ should flow towards $G|_\tau$, giving the dynamics an $\omega\to\tau$ directionality.  The most natural examples of directional graphs are those where $G$ has an explicit feedforward architecture in $G|_\omega$, for example when $G|_\omega$ is a DAG, and there are no edges from $\tau$ back to $\omega$.  In this case, it seems intuitive that the dynamics will flow along this feedforward structure in $\omega$ and end up concentrated in $\tau$.  

It turns out that any DAG decomposition of a graph $G$ immediately yields a directional partition as intuitively predicted.

\begin{lemma}\label{lemma:DAG-decomp-directional}
If $(\omega, \tau)$ is a DAG decomposition of $G$, then $G$ is directional with direction $\omega \to \tau$.  
\end{lemma}

The key to the proof of Lemma~\ref{lemma:DAG-decomp-directional} is the well known characterization of DAGs in terms of a \textit{topological ordering} of their vertices.  Recall that $G$ is a DAG if and only if there exists an ordering of the vertices such that edges in $G$ only go from lower numbered to higher numbered vertices, i.e., if $i > j$, then $i \not\to j$.

\begin{proof}
To show that $G$ is directional, we must show that any $\sigma \subseteq V_G$ that intersects $\omega$ dies by graphical domination.  Suppose $\sigma \cap \omega \neq \emptyset$, and let $j$ be the lowest numbered vertex in $\sigma \cap \omega$ with respect to some topological ordering of the DAG $G|_\omega$.  Since all the sinks in $G$ are contained in $\tau$, $j \in \omega $ must have at least one outgoing edge in $G$, so $j \to k$ for some $k \in V_G$.  
Moreover, $j$ has no incoming edges in $G|_{\sigma\cup k}$ because of its numbering in the topological ordering.
Thus, $j$ is a proper source in $G|_{\sigma \cup k}$ and by Lemma~\ref{lemma:sources}, $\sigma \notin \FP(G)$ because $k >_\sigma j$. 
Thus every $\sigma$ with $\sigma \cap \omega \neq \emptyset$ dies by graphical domination, and so $G$ is directional with direction $\omega \to \tau$.  
\end{proof}

\begin{figure}[!ht]
\vspace{-.1in}
\begin{center}
\includegraphics[width=.75\textwidth]{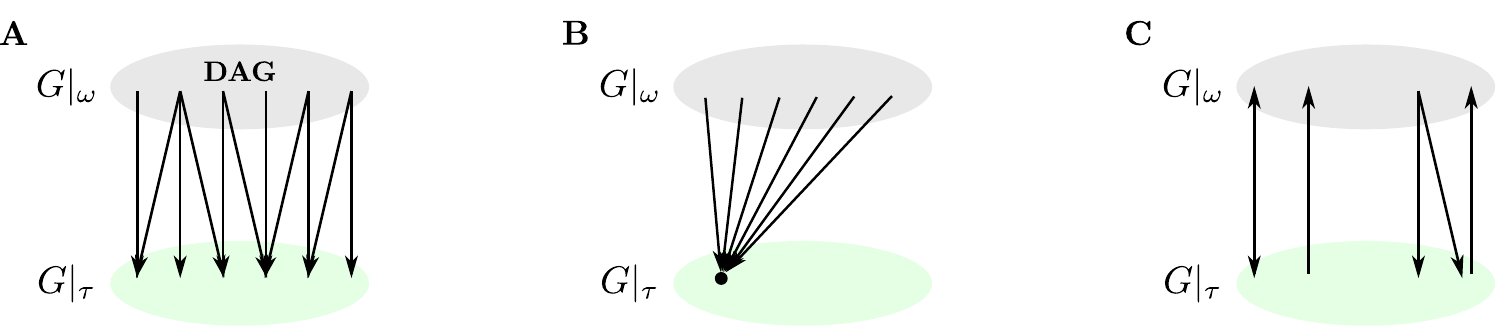}
\vspace{-.05in}
\caption{{\bf Three types of directional graphs.} 
(A) A nontrivial DAG decomposition $(\omega,\tau)$ is a directional graph with direction $\omega \to \tau$. 
(B) If $\tau$ contains a target node of $\omega$, and there are no back edges $\tau \to \omega$, then $G$ is directional irrespective of the structure of $G|_\omega$. 
(C) A more general directional graph can have a variety of forward and backward edges.}
\label{fig:dir-graphs-cartoon}
\end{center}
\vspace{-.25in}
\end{figure}

DAG decompositions are a very special case of directional graphs $\omega \to \tau$ where there are no back edges from $\tau$ to $\omega$, and the $\omega$ component of the graph is a DAG. Neither condition needs to hold for more general directional graphs.
Figure~\ref{fig:dir-graphs-cartoon} shows several types of directional graphs. In panel A, there are only edges from $\omega \to \tau$ as in a DAG decomposition. In panel B, the existence of a target in $\tau$ that receives edges from all nodes in $\omega$ guarantees that $G$ is directional irrespective of the structure of $G|_\omega$. Finally, in panel C we see a schematic of a directional graph with both forward edges from $\omega$ to $\tau$ and backward edges from $\tau$ to $\omega$.

\begin{figure}[!ht]
\begin{center}
\includegraphics[width=.9\textwidth]{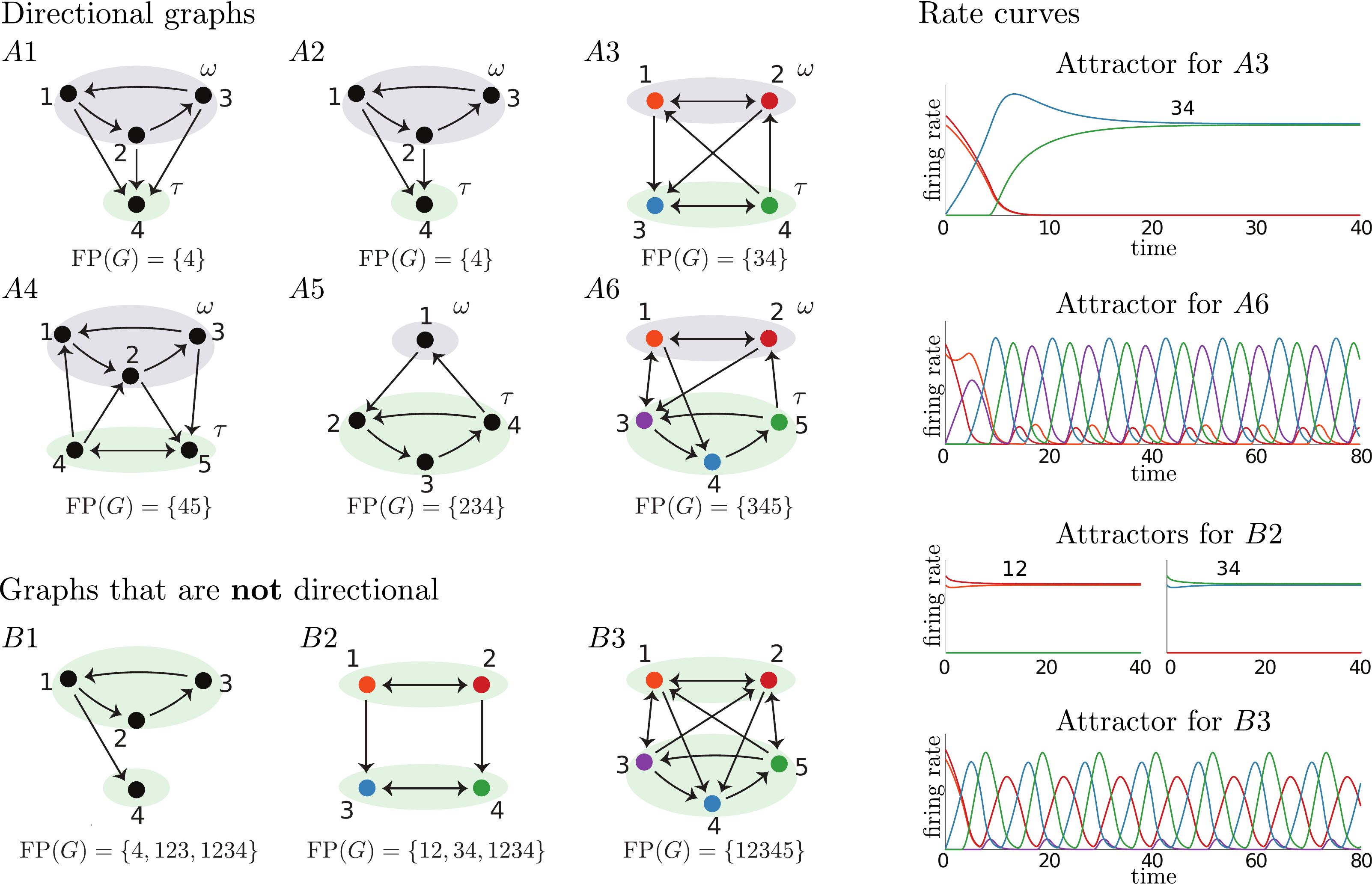}
\vspace{-.05in}
\caption{{\bf Directional graphs: examples and non-examples.} Note that we refer to fixed point supports $\{i_i,i_2,\ldots,i_k\}$ simply as $i_1i_2\cdots i_k$. For example, $234$ denotes $\{2,3,4\}$.}
\label{fig:directional-graphs}
\end{center}
\end{figure}

In fact, directional graphs can have a surprisingly large number of back edges while still preserving their ``forward'' directionality. All the graphs in Figure~\ref{fig:directional-graphs}A are directional with $\omega \to \tau$, and each of the graphs in A3 -- A6 actually has as many back edges from $\tau$ to $\omega$ as they do forward edges.  The dynamics for A3 and A6 are shown on the right, and we see that even if we initialize the activity purely on nodes in $\omega$, the activity flows $\omega \to \tau$ as predicted by the directionality.  Note that none of the graphs in panel A has a proper source, and thus none has a nontrivial DAG decomposition. 

Panel B in Figure~\ref{fig:directional-graphs} shows some example graphs that are {\it not} directional for any partition of the vertices. This is because every vertex is involved in at least one fixed point support, so there cannot be a collapse $\FP(G) \subseteq \FP(G|_\tau)$ for any $\tau \subsetneq V_G$.  The graph in B2 is particularly surprising since it only has edges forward from the 2-clique $12$ to $34$, so we might expect this to yield a directional decomposition.  But the forward edges are not sufficient to kill the 2-clique $12$, and so we see from the dynamics on the right, that we are not guaranteed a directionality of flow.  Instead, $12$ supports a stable fixed point, and thus when we initialize activity on those nodes, it remains there, and never flows to the other stable fixed point $34$.

\begin{remark}
If $G$ is a directional graph, its directional decomposition is not unique. 
For example, as long as $\omega$ has more than one vertex, then vertices can always be moved from it to $\tau$ and maintain directionality.  
However, directional decompositions are most useful when the $\tau$ component is as small as possible, since this gives the strongest restrictions on the possible fixed point supports of the whole network.   
One candidate $\tau$ for such a decomposition is $\tau := \cup_{\sigma \in \FP(G)} \sigma$.  However, this set does not guarantee a directional decomposition since we have not guaranteed that all subsets of $V_G$ that intersect $\omega := V_G \setminus \tau$ die by graphical domination. In order to satisfy this property, it may be necessary to add some additional vertices to $\tau$, and doing this in a minimal way may not be unique.  It is an open question if every directional graph has a unique directional decomposition with minimal $\tau$.  
\end{remark}

\section{Directional covers and nerve theorems}\label{sec:dir-covers}
In this section, we aim to characterize the fixed points of more complex graphs by covering the graph with directional graphs, and then analyzing a simpler associated object known as the \textit{nerve} of the cover. 
The intuition is as follows.  As described in Section~\ref{sec:dir-graphs}, if $G$ is a directional graph with direction $\omega \to \tau$, the activity of the network flows from $\omega$ to $\tau$.  Thus, from a bird's eye view, the flow of activity of such a graph can be represented by the flow of activity along a single directed edge from source to sink.  Moreover, this flow of activity reflects restrictions imposed on the fixed point supports as well.  With this in mind, we will take any graph $G$ and aim to cover it with directional graphs that have appropriate pairwise intersections.  From this cover, we construct a nerve, which is a simplified graph where subsets of vertices are collapsed to single points, and each directional graph of the cover is now represented by a single directed edge.  These edges are glued to one another in a way representative of the intersection pattern of the cover.  We will see that, with this construction, we are able to deduce certain restrictions on the fixed point supports of the original graph $G$ by studying the fixed points of the nerve of the cover, which is in general a simpler graph.

\subsection{Directional covers and nerves}
We begin by making the notion of directional cover and its nerve precise.

\begin{definition}[graph cover]
Let $G$ be a graph.  A \textit{graph cover} of $G$ is a collection of induced subgraphs $\U=\{ G_i := G|_{V_i} ~|~ \text{for some } V_i \subseteq V_G\}$ such that $G$ is the union of the $G_i$.  In other words, $V_G = \cup_{i \in I} V_i$ and $E_G = \cup_{i \in I} E_{G_i}$.  
\end{definition}
\begin{remark}
Note that every vertex and every edge of $G$ must live in at least one $G_i$, but often they live in multiple $G_i$ within the cover.  In particular, since the covering graphs are induced subgraphs of $G$, if $u, v \in V_i$ and $u, v \in V_j$, then any edges between $u$ and $v$ will be in both $G_i$ and $G_j$.  
\end{remark}

Next we turn to a special type of graph cover which we call \textit{rigid directional cover}.  In a rigid directional cover, we require that all the graphs of the cover are directional and that they overlap in prescribed ways that will facilitate associating a nerve to the cover and ensure that this nerve captures constraints on $\FP(G)$.  

\begin{figure}[!h]
  \begin{center}
  \includegraphics[width=6in]{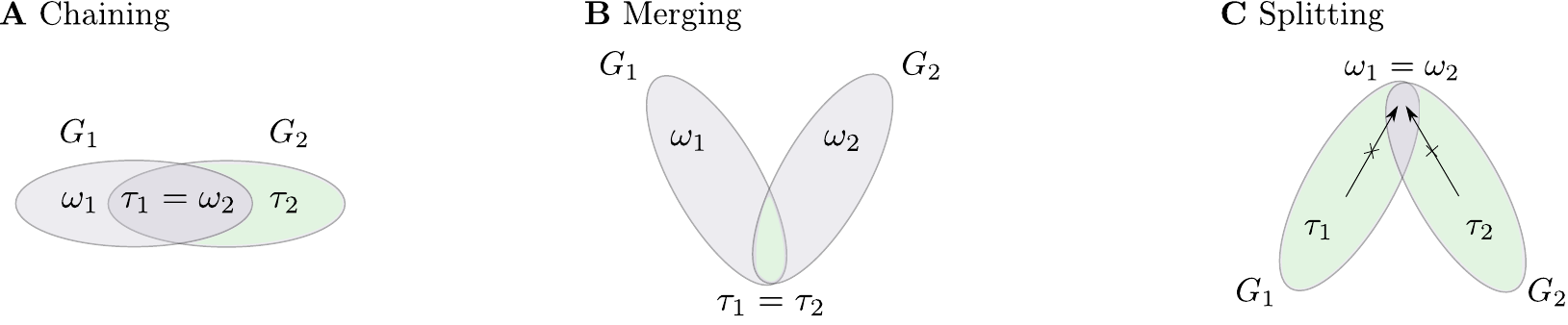}
  \caption{A pair of graphs $G_1$ and $G_2$ that have (A) a chaining overlap (B) a merging overlap and (C) a splitting overlap.}
    \label{fig:graph-overlap}
  \end{center}
\end{figure}

The rigid condition can be informally described as follows.  Consider a graph cover $\U$ of $G$, where all the covering graphs are directional. Let $G_1,G_2\in\U$ be a pair of graphs in the cover,  with directional decompositions $\omega_1 \to \tau_1$ and $\omega_2 \to \tau_2$, respectively.  The graph cover $\U$ is {\it rigid} if for any pair $G_1$ and $G_2$ that have nontrivial intersection, their overlap is of one of the following three types:
\begin{enumerate}
\item The $\tau$ component of the first graph acts as the $\omega$ component of the second, i.e., $V_{G_1} \cap V_{G_2} = \tau_1=\omega_2$.  
In this case we say the graphs have a \textit{chaining overlap}.  (See Figure~\ref{fig:graph-overlap}A.)

\item The two covering graphs intersect exactly at their $\tau$ component, i.e., $V_{G_1}\cap V_{G_2} = \tau_1 = \tau_2$. 
 In this case we say the graphs have \textit{merging overlap}. (See Figure~\ref{fig:graph-overlap}B.)
 
\item  The two covering graphs intersect exactly at their $\omega$ component,  i.e.,  $V_{G_1}\cap V_{G_2} = \omega_1=\omega_2$,
and have the additional property that there are no back edges from vertices  in $\tau$ to vertices  in $\omega$ in either graph, i.e., $E_{G_1}(\tau_1,\omega_1)=E_{G_2}(\tau_2,\omega_2)=\emptyset$.
In this case we say the graphs have a \textit{splitting overlap}. (See Figure~\ref{fig:graph-overlap}C.)
\end{enumerate}
Effectively, a rigid directional cover is always induced by a partition of the vertices of the underlying graph and this partition encodes all the information of the cover itself.  Therefore, we formally define a rigid graph cover as follows.

\begin{definition}[directional cover and its nerve]
\label{def:directional_cover}
Let $G$ be a graph.  Given a partition of the vertices, $\nu = \{\nu_1, \ldots, \nu_n\}$, let
\[E = E(G, \nu) := \{ (i,j) \in [n] \times [n]~|~ G|_{\nu_i \cup \nu_j} \text{ is directional with direction } \nu_i \to \nu_j\},\]
\[\mathcal{I}=\mathcal{I}(G,\nu):=\{ i\in [n] ~|~ G|_{\nu_i} \text{ is disconnected from the rest of the graph}\}.\]
We say that the partition $\lbrace \nu_1, \ldots, \nu_n\rbrace $ induces a \textit{rigid directional cover} of $G$ if: 
\begin{enumerate}[itemsep=.07in]
\item For every pair $(\nu_i, \nu_j)$ either $(i,j)\in E$ or $(j,i) \in E$ or there are no edges between $\nu_i$ and $\nu_j$.  In other words, the set $\U=\{G_{ij}:= G|_{\nu_i \cup \nu_j} ~|~ (i,j) \in E\}\cup\lbrace G|_{\nu_i} \mid  i\in \mathcal{I}\rbrace$ is a graph cover of $G$.

\item Whenever $G_{ij}, G_{ik}\in \U$, they have ``splitting overlap", meaning there are no edges from $\nu_j$ to $\nu_i$ and no edges from $\nu_k$ to $\nu_i$, i.e., $E_{G}(\nu_j, \nu_i)=E_{G}(\nu_k,\nu_i)=\emptyset$.
\end{enumerate}

We define the \textit{nerve} of the cover, denoted by $\N=\N(G,\U)$, to be the graph with vertex set $V_\N:=[n]$ and edge set $E_\N:=E$.  
The partition $\nu$ induces a canonical quotient map $\pi:V_G \to V_\N$ that identifies all the vertices of a component $\nu_i$, so that $\pi(\nu_i) = \{i\}$ for each $i \in V_\N$.
\end{definition}

Note that an arbitrary partition will not typically induce a rigid directional cover because there will be pairs \mbox{$(\nu_i$, $\nu_j)$} with edges between them, but the induced subgraph $G|_{\nu_i\cup\nu_j}$ will not be directional. In contrast, partitions that do induce a rigid directional cover must have $G|_{\nu_i\cup\nu_j}$ directional whenever there are edges between $\nu_i$ and $\nu_j$.  Note that we do not require the $G|_{\nu_i}$ for $i \in\mathcal{I}$ to be directional; these graphs are included in $\U$ simply to ensure that isolated components of the graph are still covered.
In this paper we will only work with rigid directional covers.  Therefore, in the remainder of this work we will use the term {\it directional cover} to refer to a rigid directional cover.

\begin{figure}[!ht]
  \centering
  \includegraphics[width=6in]{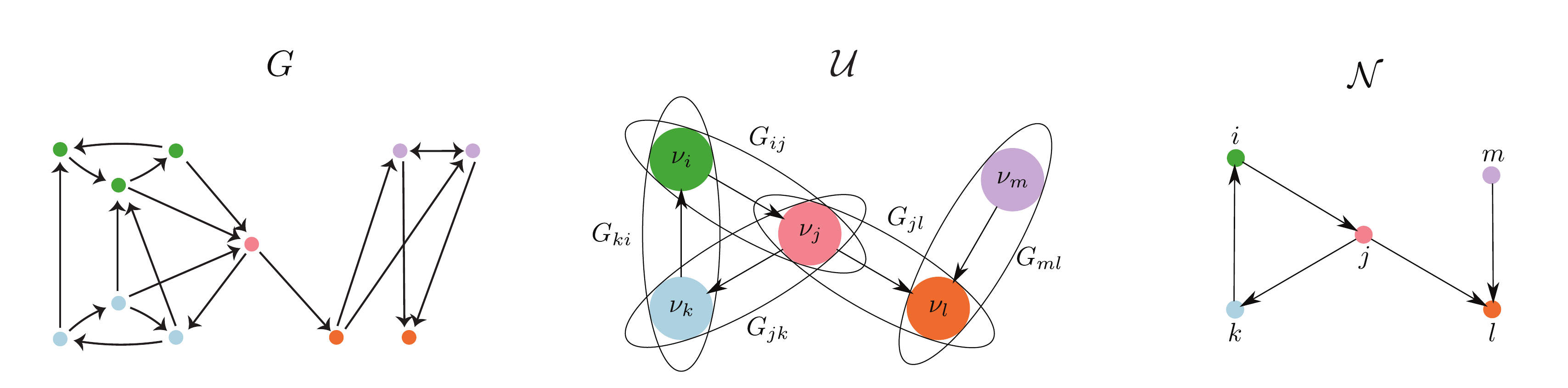}
  \caption{{\bf A graph with a directional cover and the corresponding nerve.} (Left) A graph $G$.  (Middle) A directional cover $\U$ of $G$.  The sets $\nu_i, \nu_j, \nu_k, \nu_l, \nu_m$ partition the vertices of $G$.  All the edges of $G$ are contained within some of the covering graphs $\{G_{ij}\}$ shown, each of which is directional with direction $\nu_i \to \nu_j$.  (Right) The nerve $\N=\N(G,\U)$ with a vertex for each component of the partition and a directed edge for each directional graph of the cover. 
  } 
  \label{fig:cartoon-dir-cover}
\end{figure}

Figure~\ref{fig:cartoon-dir-cover} gives an illustration of a graph $G$ with vertex partition $\nu=\{\nu_i\}$ that induces a collection of covering graphs $\{G_{ij}\}$ that are all directional.  In the nerve, $\N=\N(G,\U)$, we see a vertex $i$ for each component $\nu_i$ from $G$, and an edge $i \to j$ corresponding to each covering graph $G_{ij}$, which has direction $\nu_i \to \nu_j$.    

\begin{lemma}
 Let $G$ be a graph with nerve $\N=\N(G,\U)$ for some directional cover $\U$.  Then the nerve $\N$ is a simple directed graph that is oriented.
\end{lemma}
\begin{proof}
Recall that given a partition $\nu=\{\nu_1, \ldots, \nu_n\}$ of the vertices of $G$, the nerve $\N$ is defined as a graph on $n$ vertices with edge set $E$ given in Definition~\ref{def:directional_cover}.  The edge set $E$ is straightforward to determine from the partition $\nu$.  Specifically, if there are no edges between $\nu_i$ and $\nu_j$ in $G$, then neither $(i,j)$ nor $(j,i)$ are in $E$, since $G|_{\nu_i \cup \nu_j}$ is a disjoint union, which can never be directional.  If there are any edges between $\nu_i$ and $\nu_j$, we must have either $(i,j) \in E$ or $(j,i) \in E$ in order for the $\{G_{ij}\}$ to cover $G$.  Moreover, we can only have one of $(i,j)$ or $(j,i)$ in $E$ since $G|_{\nu_i \cup \nu_j}$ can never be directional with both $\nu_i \to \nu_j$ and $\nu_j \to \nu_i$.  Thus, $\N$ is oriented.
Finally, to see that $\N$ is simple, notice that $(i,i) \notin E$ since $G_{ii}$ can never be directional with $\omega=\tau=\nu_i$.  
\end{proof}

Given the complexity of the requirements of a directional cover, specifically that every covering graph of the form $G|_{\vu_i \cup \vu_j}$ be directional, it is natural to ask when a graph actually has such a cover.
Of course, every graph has a trivial directional cover induced by the trivial partition $\nu_1 = V_G$; in this case, the nerve of the cover is just a single point.  At the other extreme, whenever $G$ is an \emph{oriented} graph, the partition of singletons $\nu_i=\{i\}$ will induce a directional cover, whose nerve is precisely the original graph $G$.  While these two trivial covers exist, they clearly do not provide any insight into the structure or expected dynamics of $G$.  There is an art to finding a partition of $V_G$ from which a cover with an informative nerve can be obtained.

It is important to note that not every graph has a directional cover induced by a nontrivial partition.  For example, if $G$ is a clique, then there is no nontrivial partition of $V_G$ that can admit a directional cover since every $G_{ij}$ will be a clique, and thus not directional.  
At the other extreme, there are graphs with multiple nontrivial partitions of $V_G$ which induce a directional cover.  See for example, Figure~\ref{fig:counterexample}C-D, which shows two different covers of the same graph, and where the nerve of the first one is directional while the nerve of the second is a cycle.  

It is an open question which graphs have at least one nontrivial partition that admits a directional cover.  And unfortunately, there is currently no efficient way to find all the partitions of a graph that do induce directional covers.  
However, when we have a directional cover of $G$, obtained either by brute force search or from intuition into the original construction of the graph, the nerve of the cover can give significant insight into the collection of fixed points and consequently into the predicted dynamics of the underlying network. In particular, nerve theorems ensure that there is a provable connection between $\FP(G)$ and the structure of the nerve under certain conditions on $G$ and/or $\N(G, \U)$.

\subsection{Nerve theorems}
Ideally, we would hope for a nerve theorem that provides a strong connection between the fixed point supports of the original graph and those of the nerve.  For example, we might hope that for any graph $G$ that admits a directional cover with nerve $\N$ that we can guarantee a condition such as 
\begin{equation}\label{eq:ideal-nerve-thm}
\sigma \in \FP(G) \ \Rightarrow \ \pi(\sigma) \in \FP(\N).
\end{equation}
Unfortunately, though, this strong restriction on $\FP(G)$ does not hold for all graphs and all directional covers.

\begin{figure}[!ht]
  \centering
  \includegraphics[width=120mm]{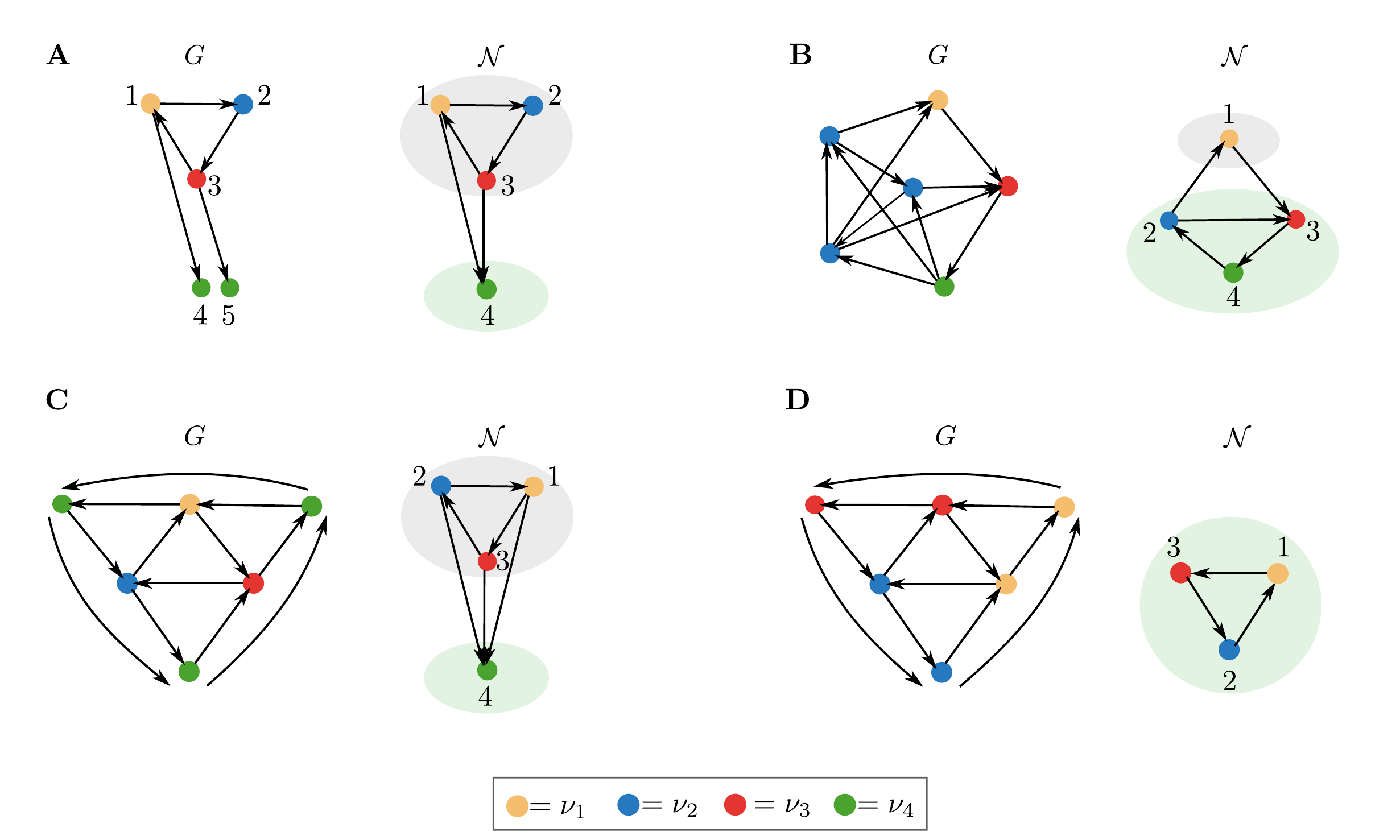}
  \caption{{\bf Counterexamples to general nerve theorems.} (A) and (B) give two different graphs with partitions that induce directional covers.  The nerve of each cover is shown to the right.  (C) and (D) give two different partitions for the \textit{same} graph.  In C, the partition induces a nerve that is directional, while in D, the partition induces a nerve that is a cycle.}

  \label{fig:counterexample}
  \vspace{-0.5cm}
\end{figure}

\begin{example}\label{ex:dir-nerve1}
For the graph in Figure~\ref{fig:counterexample}A, we see that the partition $\lbrace \nu_1, \ldots, \nu_4\rbrace$ shown induces a directional cover of $G$. The nerve $\N=\N(G,\U)$ of this cover is shown on the right.  
Recall that a cycle is \textit{uniform in-degree 1} and thus it supports a fixed point precisely when no external vertex receives more than one edge from it (see Theorem~\ref{thm:uniform-in-degree} in Section~\ref{sec:background} ).  Thus we have $123 \in \FP(G)$\footnote{
Recall that we write 123 to denote the fixed point support $\{1,2,3\}$.}
since the external vertices  4 and 5 each receive only one edge from the cycle.  However, $\pi(123) = 123 \notin \FP(\N)$ since in the nerve, the cycle 123 has two outgoing edges to vertex 4.  Thus, $\sigma \in \FP(G) \not\Rightarrow \pi(\sigma) \in \FP(\N)$ for this example graph.
\end{example}

An alternative style of nerve theorem would enable us to at least restrict the candidate fixed point supports of $G$ based on the structure of the nerve.  For example, we might hope that whenever the nerve $\N(G, \U)$ is directional with direction $\W \to \T$ that the directionality would pullback to guarantee $G$ is directional with direction $\omega \to \tau$ for $\omega =\pi^{-1}(\W)$ and $\tau=\pi^{-1}(\T)$. If this held, then we could guarantee that the fixed point supports of $G$ were confined to $\tau=\pi^{-1}(\T)$, and thus $\FP(G) \subseteq \FP(G|_\tau)$ by Lemma~\ref{lemma:cor2}.  Such a result would be somewhat weaker than~\eqref{eq:ideal-nerve-thm} in that the restrictions on $\sigma \in \FP(G)$ would not be as strong.  
On other hand, it would also be somewhat stronger in a different direction, since a result like this would guarantee the presence of graphical domination relationships for ruling out fixed point supports, which is not something guaranteed by~\eqref{eq:ideal-nerve-thm}.
Unfortunately, this alternative nerve theorem does not hold in general either, and the same graph as above provides a counterexample, as do the other graphs in Figure~\ref{fig:counterexample}.

\begin{example}
It is straightforward to see that the nerve $\N$ in Figure~\ref{fig:counterexample}A is directional for $\W:= \{1,2,3\}$ and $\T:= \{4\}$: every subset $\S \subseteq V_\N$ that intersects $\W$ either has a proper source in $G|_\S$, and thus dies from domination by Lemma~\ref{lemma:sources}, or contains $123$, in which case we have $4>_\S 1$.  However, $G$ is not directional since $12345 \in \FP(G)$, so there is no collapse of the fixed point supports of $G$ onto a proper subset $\tau$. 
Figure~\ref{fig:counterexample}C gives another counterexample where $G$ has a full support fixed point but a directional cover whose nerve is directional. 
\end{example}

We have seen that in general the existence of a directional relationship $\W \to \T$ of the nerve does not guarantee directionality of $G$.  But are there certain conditions under which this holds?  It turns out that we can pullback such a directionality relationship in the special case when the nerve has a nontrivial DAG decomposition (see Definition~\ref{def:DAG-decomp}).  

\begin{theorem}[DAG decomposition of the nerve]\label{thm:DAG-decomp}
 Let $G$ be a graph with nerve $\N=\N(G,\U)$ where $\U$ is a directional cover induced by a partition $\lbrace\nu_1, \ldots, \nu_n\rbrace$, and let $\pi:V_G\to V_\N=[n]$ be the canonical quotient map of the partition.
Then for any DAG decomposition $(\W, \T)$ of the nerve $\N$, we have that $G$ is directional with direction $\omega \to \tau$ for $\omega=\pi^{-1}(\W)$ and $ \tau = \pi^{-1}(\T)$.
In particular, 
\[\FP(G) \subseteq \FP(G|_{\tau}),\]
and so for all $\sigma \in \FP(G)$, we have $\pi(\sigma)\subseteq \T$.
\end{theorem}

Notice that in Theorem~\ref{thm:DAG-decomp}, we can only conclude that $\sigma \in \FP(G) \ \Rightarrow \ \pi(\sigma)\subseteq \T$, and so we do not quite have the ideal nerve theorem result that $\pi(\sigma) \in \FP(\N)$ as in~\eqref{eq:ideal-nerve-thm}.  The conclusion in Theorem~\ref{thm:DAG-decomp} is weaker than~\eqref{eq:ideal-nerve-thm} because although the directionality of $\N$ guarantees that every element of $\FP(\N)$ is contained in $\T$ as is each $\pi(\sigma)$, we cannot guarantee that $\pi(\sigma)$ is actually a fixed point support of $\N$.

Next we consider when the nerve $\N$ is itself a DAG so that, in the maximal DAG decomposition, $\T$ is precisely the sinks of $\N$.  We can immediately apply Theorem~\ref{thm:DAG-decomp} to see that the directionality of $\N$ pulls back to $G$, but in fact we can say something stronger: the ideal nerve theorem conditions of~\eqref{eq:ideal-nerve-thm} hold in this case.  Moreover, it turns out that there are further restrictions on the fixed point supports of $G$ in terms of the fixed points of the component subgraphs $G|_{\nu_i}$, which are prescribed by the partition.  

\begin{theorem}[DAG nerve]\label{thm:DAG}
 Let $G$ be a graph with nerve $\N=\N(G,\U)$ where $\U$ is a directional cover induced by a partition $\{\nu_1, \ldots, \nu_n\}$, and let $\pi:V_G\to V_\N=[n]$ be the canonical quotient map of the partition.
 Suppose that $\N$ is a DAG, and let $\T = \{\text{sinks of } \N\}$ and $\W=V_\N \setminus \T$.  Then $G$ is directional with direction $\omega \to \tau$ for $\omega=\pi^{-1}(\W)$ and $ \tau = \pi^{-1}(\T)$.
 
 Moreover,
\begin{enumerate}
\item $ \sigma \in \FP(G) \ \Rightarrow \ \pi(\sigma) \in \FP(\N) = \P(\T)\setminus\{\emptyset\}$, where $\P(\T)$ denotes the power set of $\T$.  
\item $ \sigma \in \FP(G) \ \Rightarrow \ \sigma \cap \nu_i \in \FP(G|_{\nu_i}) \cup \{\emptyset\}  \text{ for all } i \in \T$ and  $\sigma \cap \nu_j = \emptyset  \text{ for all } j \in \W$.
\end{enumerate}
 \end{theorem}

\begin{figure}[!ht]
  \begin{center}
  \includegraphics[width=6.5in]{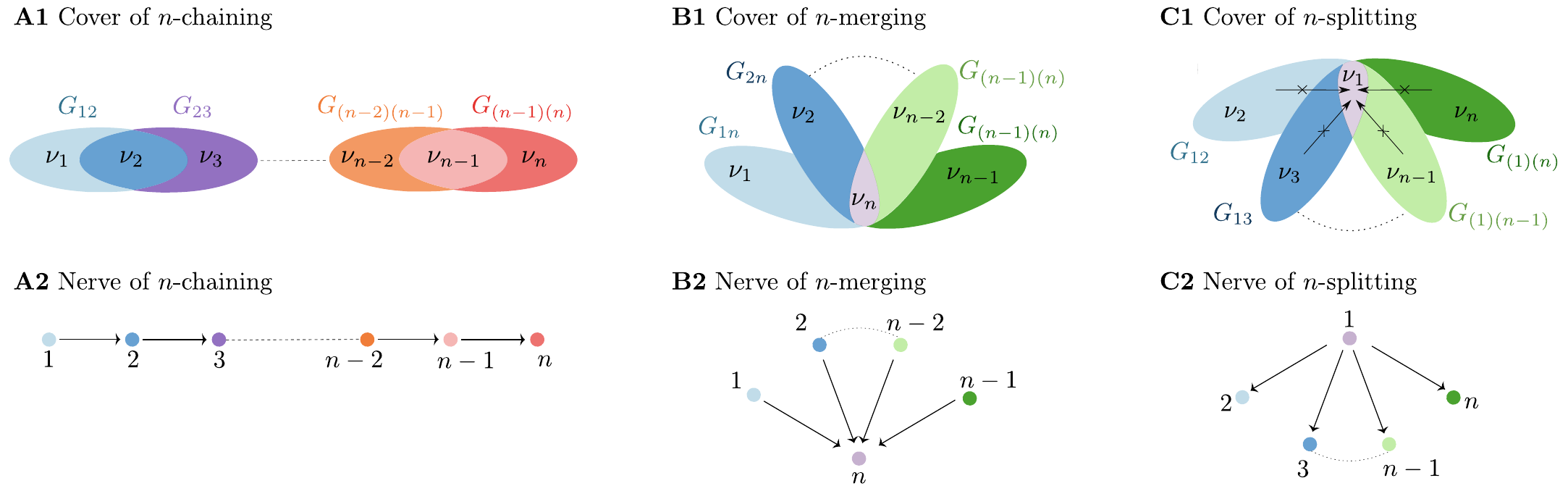}
  \vspace*{-0.5cm}
  \caption{{\bf Example directional covers and nerves.}  (A1-C1) Graphs with simple directional covers in which every pair of covering graphs have the same type of overlap (chaining overlap in A, merging overlap in B, and splitting overlap in C). (A2-C2) Nerves for the simple directional covers above.}
    \label{fig:nerve-multipleoverlaps}
  \end{center}
\end{figure}

Figure~\ref{fig:nerve-multipleoverlaps} illustrates three special cases of directional covers $\U$ of a graph $G$ with their corresponding nerves shown below.  
These directional covers have pairwise overlaps that are either:
only chainings (Figure~\ref{fig:nerve-multipleoverlaps}A), only mergings (Figure~\ref{fig:nerve-multipleoverlaps}B), or only splittings (Figure~\ref{fig:nerve-multipleoverlaps}C). We refer to these types of overlaps as $n$-chaining, $n$-merging and $n$-splitting, respectively.  
We see that their corresponding nerves are DAGs where the set of sinks $\T$ is either a single sink, $\T=\{ n\}$, as in panels A2 and B2, or an independent set of sinks, $\T=\lbrace 2,3,\ldots,n\rbrace$, as in panel C2.  
Theorem~\ref{thm:DAG-decomp} tells us that in each case the underlying graph $G$ is directional with direction $\omega \to \tau$ for $\tau = \pi^{-1}(\T)$.  
Additionally, in the case of $n$-splitting, Theorem~\ref{thm:DAG} gives a stronger result.  Namely, 
$$\sigma \in \FP(G) \ \Rightarrow \ \pi(\sigma) \in \FP(\N) \ \text{ and } \ \sigma\cap\nu_i \in\FP(G|_{\nu_i})\cup \lbrace \emptyset \rbrace \ \text{ for } i\in\lbrace 2, 3, \ldots, n\rbrace.$$
That is: any fixed point support $\sigma$ of $G$ gets pushed forward to a fixed point support of the nerve $\N$, 
 and for any $2\leq i \leq n$ we have that if $\sigma$ intersects $\nu_i$ then this intersection is also a fixed point support of the induced subgraph on $\nu_i$.

Thus far, we have only seen nerve theorems in the case when $\N$ has a nontrivial DAG component, but it turns out that a similar nerve result holds in the case when $\N$ is a cycle (and thus has no DAG component).   

\begin{theorem}[cycle nerve]\label{thm:cycle-nerve}
 Let $G$ be a graph with nerve $\N=\N(G,\U)$ where $\U$ is a directional cover induced by a partition $\lbrace\nu_1, \ldots, \nu_n\rbrace$, and let $\pi:V_G\to V_\N=[n]$ be the canonical quotient map.\\
 Suppose that $\N$ is a cycle on $n$ vertices. Then 
\begin{enumerate}
\item $ \sigma \in \FP(G) \ \Rightarrow \ \pi(\sigma) \in \FP(\N) = \{[n]\}$
\item If  $\lbrace \nu_1, \ldots, \nu_n\rbrace$ is a \textit{simply-added partition} of $G$, then 
$\sigma \in \FP(G) \ \Rightarrow \ \sigma \cap \nu_i \in \FP(G|_{\nu_i})  \text{ for all } i \in [n].$
\end{enumerate}
 \end{theorem}

Theorem~\ref{thm:cycle-nerve} is a repackaging of results from \cite{graph-rules-paper}, which explores graphs known as \textit{directional cycles}; in the terminology of this paper, these are precisely graphs with a directional cover whose nerve is a cycle.  In \cite[Theorem 1.2]{graph-rules-paper}, it was shown that for this family of graphs, every fixed point support must nontrivially intersect every $\nu_i$, and so for every $\sigma \in \FP(G)$, we have $\pi(\sigma) = [n]$.  Recall that when $\N$ is a cycle, $\FP(\N) = \{[n]\}$ by Lemma~\ref{lemma:FP-cycle}, and thus, combining these results, we are guaranteed that $\pi(\sigma) \in \FP(\N)$.  In \cite[Theorem 1.5]{graph-rules-paper}, it was also shown that when the partition $\lbrace\nu_1, \ldots, \nu_n\rbrace$ has a special property, known as \textit{simply-added}\footnote{We say that $\{\nu_1, \ldots, \nu_n\}$ is a \textit{simply-added partition} if every vertex in $\nu_i$ is treated identically by the rest of the graph.  In other words, for every $j \in V_G \setminus \nu_i$ if $j \to k$ for some $k \in \nu_i$, then $j \to \ell$ for all $\ell \in \nu_i$.}, then every fixed point support must restrict to a fixed point in each of the component subgraphs $G|_{\nu_i}$, yielding the second part of Theorem~\ref{thm:cycle-nerve}.  

It is worth noting that another special family of graphs with directional covers was previously studied in 
\cite[Section 5]{fp-paper}. That work focused on \textit{composite graphs}, which are graphs where all the vertices  in a component behave identically with respect to the rest of the graph.  Consequently, the only directional covering graphs $G_{ij}$ used in the cover are those that have all possible edges forward from $\nu_i$ to $\nu_j$ and no backward edges.  In this context, the components of a composite graph correspond to the partition $\lbrace\nu_1, \ldots, \nu_n\rbrace$ that induces the directional cover, and the \textit{skeleton} of the composite graph is its nerve.  With this perspective, many of the results of \cite[Section 5]{fp-paper} can be reinterpreted as nerve theorems for the special family of composite graphs.

\subsection{Proofs of nerve theorems}\label{sec:nerve-proofs}

Throughout this subsection we fix the following notation: $G$ is a graph with a partition $\lbrace \nu_1, \ldots, \nu_n\rbrace$ that induces a directional cover $\U$. The nerve $\N(G,\U)$ is denoted by $\N$, and $\pi:V_G\to V_{\N}=[n]$ is the canonical quotient map induced by the partition.

Before proving the nerve theorems we give an overview of the structure of the proofs.  
To prove Theorem~\ref{thm:DAG-decomp} (DAG decomposition of the nerve), we first show that whenever $\N$ has a proper source $s$, we can guarantee that $G$ is directional for $\omega = \pi^{-1}(s)$ and $\tau=V_G \setminus \omega$ (see Lemma~\ref{lemma:proper_source_nerve}).
This gives a rather coarse directional decomposition of $G$.  
We will then consider the general case when we have a DAG decomposition $(\W,\T)$ of the nerve $\N$.  We will use the previous result and show inductively that $G$ is directional with direction $\pi^{-1}(\W) \to \pi^{-1}(\T)$.
For this proof, we will use three ingredients we briefly describe now.  

First, we will use a \textit{topological ordering} on $\W$, which guarantees that the only possible edges in $\N|_\W$ are from lower numbered vertices to higher number vertices.  With respect to this topological ordering, vertex 1 in $\N$ is a proper source; vertex 2 is a proper source in $\N|_{V_\N \setminus \{1\}}$; vertex 3 is a proper source in $\N|_{V_\N \setminus \{1, 2\}}$, and so on.  This ordering will allow us to induct on $| \W|$.
The second ingredient is Lemma \ref{lemma:dir-iterate}.  This result will allow us to  refine a directional decomposition of a graph in order to grow $\omega$, and consequently shrink $\tau$ by looking at directional decompositions of the subgraph induced on the vertices in $\tau$.  
The third and final ingredient is  Lemma~\ref{lemma:restricted-nerve}.  This result will show that the construction of a directional cover and its nerve is compatible with taking subgraphs of the underlying graph corresponding to only some of the components of the partition inducing the cover.

At the core of the proofs is the process of ``extending domination".  To see why this idea is central, consider a directional cover of $G$ and  $\sigma\subseteq V_G$ a subset of the vertices of $G$.  
Let $\mu$ be the intersection of $\sigma$ with the vertices of one of the covering graphs such that $\mu$ is not contained in the $\tau$ component of that covering graph.  
Since $\mu$ is completely contained within a directional graph then it must die by domination. 
Under certain conditions, one can show that $\sigma$ dies by domination by extending the domination relationship on $\mu$ to all of $\sigma$.
Therefore, all the proofs hinge on the following technical result that determines when domination can be extended to a superset.

\begin{figure}[!h]
\vspace{-.1in}
  \centering
  \includegraphics[width=1.25in]{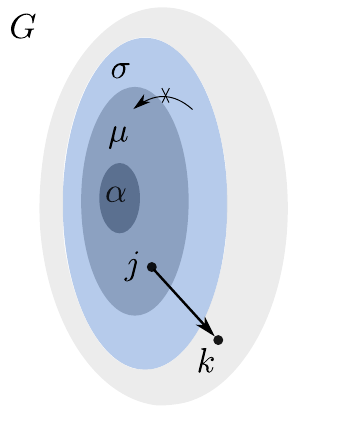}
  \vspace*{-0.3cm}
  \caption{{\bf Figure for Lemma~\ref{lemma:lemma0}.} A graph $G$ with subsets of the vertices $\alpha\subset \mu \subseteq \sigma \subset V_G$ satisfying the conditions of Lemma~\ref{lemma:lemma0}.  Specifically, the vertex $j \in \mu\setminus \alpha$, the vertex $k \in V_G$, and there are no edges from vertices in $\sigma$ to those in $\mu \setminus \alpha$.}
  \label{fig:extension}
\end{figure}

\begin{lemma}[restricting and extending domination]\label{lemma:lemma0}
Let $G$ be a graph, and let $\alpha \subset \mu \subset \sigma \subseteq V_G$ be subsets of vertices with $\alpha$ possibly empty (see Figure~\ref{fig:extension}).  Then the following hold:

\begin{enumerate}
\item[(a)] \textit{Restriction:} Suppose $k>_{\sigma}j$ where $j\in\mu$ and $k\in V_G$.  Then $k>_{\mu}j.$
\item[(b)] \textit{Extension:} Suppose $k>_{\mu}j$ where $j\in \mu \setminus \alpha$ and $k\in V_G$.   If there are no edges from $\sigma \setminus  \mu$ to $\mu \setminus  \alpha$ (i.e., $E_G(\sigma \setminus  \mu,~\mu \setminus  \alpha)~=~\emptyset$), then $k>_{\sigma}j.$
\end{enumerate}
\end{lemma}

\begin{proof}
Recall from Definition \ref{def:domination} that $k>_\sigma j$ if the following three conditions hold:
\begin{enumerate}
\item For all $i \in \sigma\setminus \{j,k\}$ if $i\rightarrow j$, then $i \rightarrow k$.
\item If $j \in \sigma$, then $j\rightarrow k$.
\item If $k \in \sigma$, then $k\nrightarrow j$.
\end{enumerate}

For (a), {\it Restriction}, we see that $k>_\sigma j$ immediately implies that $k>_\mu j$ since if condition 1 holds for all of $\sigma$, then it holds for the subset $\mu$ as well, and conditions 2 and 3 go through directly.

For (b), {\it Extension}, we see condition 2 goes through immediately since $j \in \mu \subset \sigma$.  For condition 3, observe that if $k\in \sigma$ then either $k \in \mu$ or $k \in \sigma \setminus  \mu$, and in both cases $k \nrightarrow j$ as required.  Finally, for condition 1, notice that for 
$i\in \mu$ this condition holds because $k>_{\mu}j$, while for $i\in \sigma\setminus \mu$, this condition holds trivially because $j\in \mu\setminus  \alpha$ and there are no edges from $\sigma \setminus  \mu$ to $\mu \setminus  \alpha$ by hypothesis.  Thus condition 1 holds as well, and so $k >_\sigma j$.
\end{proof}

We can now prove that it is possible to pull back directionality from the nerve $\N$ to $G$ whenever $\N$ has a proper source.

\begin{lemma}\label{lemma:proper_source_nerve}  
If $s$ is a proper source in $\N$, then $G$ is directional with $\omega = \pi^{-1}(s)$ and $\tau= V_G \setminus \omega$.
\end{lemma}
\begin{proof}
Let $s$ be a proper source in $\N$, $\omega := \pi^{-1}(s)$, and $\tau:= V_G \setminus \omega$.  To show that $G$ is directional with direction $\omega \to \tau$, consider $\sigma\subseteq V_G$ such that $\sigma\cap\omega\neq\emptyset$.  We need to show that $\sigma$ dies by domination, i.e., that there exists a $j\in \sigma\cap\omega$ and a $k\in V_G$ such that $k>_\sigma j$. 

The organization of the proof is as follows.  We first find a covering graph $G_{s1}\in\U$ such that $\omega \subseteq V_{G_{s1}}$.  Then we set $\mu\subseteq \sigma$ to be the restriction of $\sigma$ to $G_{s1}$.
Since $G_{s1}$ is directional, $\mu$ dies by domination.  
We will show that by setting $\alpha\subset \mu$ to be the intersection of $\mu$ with the $\tau$ component of $G_{s1}$, the conditions of Lemma~\ref{lemma:lemma0} hold  (see Figure~\ref{fig:proper_source_nerve}).
Thus, we can extend the domination relationship to all of $\sigma$.  

\begin{figure}[!h]
\begin{center}
\vspace{-.1in}
  \includegraphics[width=3in]{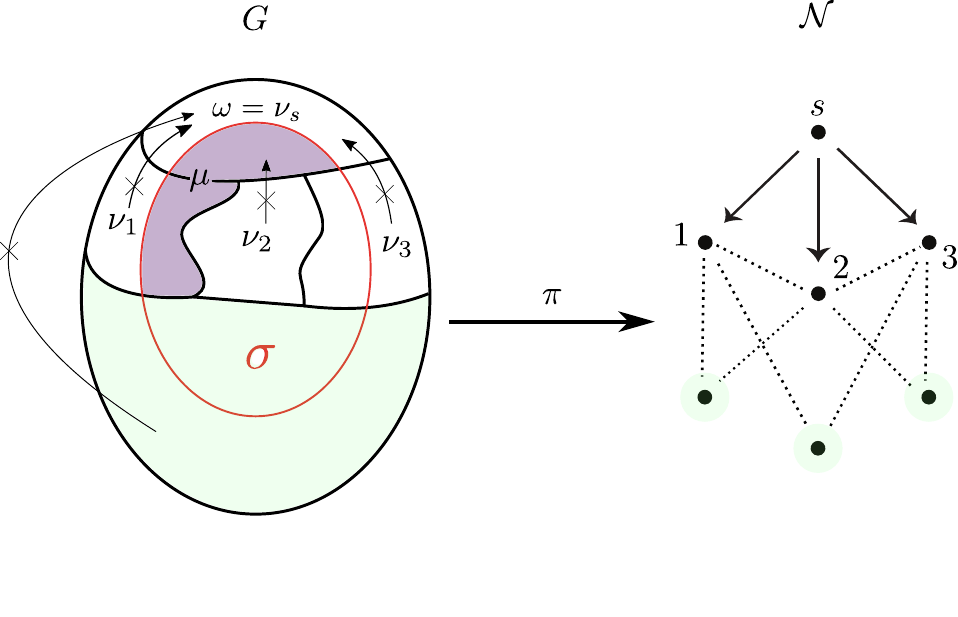}
  \vspace{-0.5cm}
  \caption{{\bf Figure for proof of Lemma~\ref{lemma:proper_source_nerve}.} Vertex $s$ is a proper source in $\N$ (right) with outgoing edges to vertices  1,2, and 3.  In $G$ (left), we see the corresponding components $\nu_s$ and $\{\nu_1, \nu_2, \nu_3\}$. We consider a subset $\sigma \subseteq V_G$ (outlined in red), and let $\mu := \sigma \cap (\nu_s \cup \nu_1)$ (shaded in purple).  Arrows with an x through them indicate that no edges are allowed in the specified direction between the relevant components. Specifically, there are no edges from $\{\nu_1, \nu_2, \nu_3\}$ into $\nu_s$ because the graphs $G_{s1}$, $G_{s2}$, and $G_{s3}$ must have ``splitting overlap" by the definition of directional cover.  There are no edges from any other vertices in $G$ into $\nu_s$ because there are no edges in the nerve between $s$ and any of the other vertices besides 1, 2, and 3. }
 \label{fig:proper_source_nerve}
\end{center}
\vspace{-.2in}
\end{figure} 

To find $G_{s1},$ notice that since $s$ is a proper source in $\N$ there exists at least one vertex in $\N$ that $s$ sends an edge to; without loss of generality, label the vertices of $\N$ that $s$ sends edges to as $1, \ldots, m$.  Since $s \to 1$ in $\N$, the covering graph $G_{s1}:= G|_{\nu_s \cup \nu_1}$ must be directional with direction $\nu_s \to \nu_1$.  Let $\mu := \sigma \cap (\nu_s \cup \nu_1)$ (see Figure~\ref{fig:proper_source_nerve}).  Since $G_{s1}$ is directional, there exists a $j \in \mu \cap \nu_s$ and $k \in V_{G_{s1}}$ such that $k >_\mu j$.  Following the notation of Lemma~\ref{lemma:lemma0}, let $\alpha:= \mu \setminus \nu_s = \mu \cap \nu_1$.  We will show that there are no edges in $G$ from vertices  in $\sigma \setminus \mu$ to vertices  in $\mu\setminus \alpha = \mu \cap \nu_s$, enabling us to extend the domination relationship from $\mu$ to all of $\sigma$.

Note that by definition of directional cover, there can only be edges between $\nu_s$ and $\nu_\ell$ in $G$ if there is an edge between $s$ and $\ell$ in $\N$ (see Definition~\ref{def:directional_cover}).  Thus the only candidate vertices in $G$ that could send edges into $\mu\setminus \alpha = \mu \cap \nu_s$ are those in $\{\nu_1, \ldots, \nu_m\}$ since the only edges in $\N$ that involve $s$ are those from $s$ to $1, \ldots, m$.   But since $(s,1), \ldots, (s,m) \in E_\N$, condition (2) of the definition of directional cover requires that the covering graphs $G_{s1}, \ldots, G_{sm}$ have splitting overlap, so there are no edges from $\nu_\ell$ to $\nu_s$ for any $1 \leq \ell \leq m$.  Thus, there are no edges from $\sigma \setminus \mu$ into $\mu \cap \nu_s = \mu \setminus \alpha$, and so by Lemma~\ref{lemma:lemma0}, the domination relationship $k>_\mu j$ extends to give $k>_\sigma j$.  Hence $G$ is directional with direction $\omega \to \tau$.  
\end{proof}

We now give the two lemmas that allow us to inductively use the result above. 
First, we show that one can refine a directional decomposition of a graph by looking at possible directional decompositions of the subgraph induced on the $\tau$ component.  

\begin{figure}[!ht]
  \centering
  \vspace{-.1in}
  \includegraphics[width=1.25in]{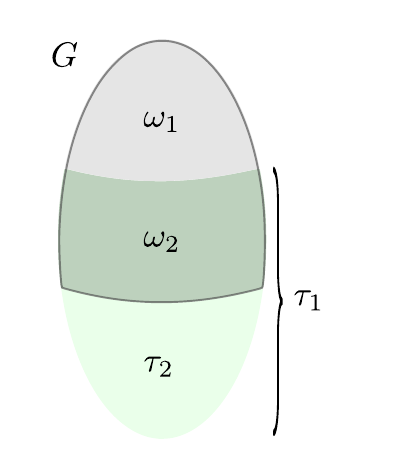}
  \caption{{\bf Figure for Lemma~\ref{lemma:dir-iterate}.} A directional graph $G$ with an initial directional partition $\omega_1 \to \tau_1$. Additionally, $G|_{\tau_1}$ is directional with direction $\omega_2 \to \tau_2$.  These two partitions can then be combined to show that $G$ is directional for a larger set $\omega = \omega_1 \cup \omega_2$ (outlined in gray) and smaller set $\tau=\tau_2$. }
  \label{fig:dir-iterate}
  \vspace{-.2in}
\end{figure}

\begin{lemma}\label{lemma:dir-iterate}
Suppose that $G$ is a directional graph with direction $\omega_1 \to \tau_1$ and that $G|_{\tau_1}$ is also directional with direction $\omega_2 \to \tau_2$ (see Figure~\ref{fig:dir-iterate}).  Then $G$ is directional with direction $\omega_1 \cup \omega_2 \to \tau_2$.  
\end{lemma}

\begin{proof}
Let $\omega = \omega_1 \cup \omega_2$ and $\tau = \tau_2$.  To show that $G$ is directional with direction $\omega \to \tau$, consider $\sigma \subseteq V_G$ such that $\sigma \cap \omega \neq \emptyset$.  We will show that $\sigma$ dies by domination.

\noindent \underline{Case 1:} $\sigma \cap\omega_1 \neq \emptyset$.  Then since $G$ is directional with $\omega_1 \to \tau_1$, $\sigma$ dies by domination.

\noindent \underline{Case 2:} $\sigma \cap\omega_1 = \emptyset$. Then $\sigma \subseteq \tau_1$.  Since $\sigma \cap \omega \neq \emptyset$, we have $\sigma \cap \omega_2 \neq \emptyset$.  Then since $G|_{\tau_1}$ is directional with $\omega_2 \to \tau_2$, there exists $j \in \sigma \cap \omega_2$ and $k \in V_{G|_{\tau_1}} = \tau_1$ such that $k>_\sigma j$ in $G|_{\tau_1}$.  Since $\sigma \cap \omega_1 = \emptyset$, there are no vertices  in $\sigma$ outside of $G|_{\tau_1}$ that could potentially subvert the domination relationship between $k$ and $j$.  Thus, $k>_\sigma j$ in all of $G$, and so $G$ is directional for $\omega = \omega_1 \cup \omega_2$ and $\tau = \tau_2$.
\end{proof}

We now show that the construction of a directional cover and its nerve behaves nicely for induced subgraphs of $G,$ restricting to a subset of the partition components.

\begin{lemma}\label{lemma:restricted-nerve} 
Let $\nu_I:=\lbrace \nu_i \mid i\in I\rbrace$ be a subset of the components of the partition $\nu$ of $G$, for $I\subseteq V_\N$. Let $G_I:=G|_{\cup_{i \in I} \nu_i}$ denote the induced subgraph of $G$ on the components $\nu_I$. Then the partition $\nu_I$ of the vertices of $G_I$ induces a directional cover $\U_I$ whose nerve is $\N(G_I,\U_I)=\N|_I,$ the restriction of the original nerve $\N = \N(G,\U)$ to the vertices $I$.
\end{lemma}

\begin{proof}
Observe that the partition $\nu_I$ yields the edge set
$$E(G_I, \nu_I) := \{ (i,j) \in I \times I~|~ G|_{\nu_i \cup \nu_j} \text{ is directional with direction } \nu_i \to \nu_j\}.$$
This edge set is clearly a subset of the edge set $E(G, \nu)$ for the cover $\U$ of $G$; specifically, $E(G_I, \nu_I) = \{(i,j) \in E(G, \nu) ~|~ i,j \in I\}$.  Moreover, the set of graphs $\{G_{ij} ~|~ (i,j) \in E(G_I, \nu_I) \}$ form a graph cover of $G_I$ because for any $i,j \in I$ if there are any edges between $\nu_i$ and $\nu_j$ in $G$, then $G_{ij}$ or $G_{ji}$ must have been in the cover $\U$ of $G$, so one of these graphs is directional, and thus $(i,j)$ or $(j, i)$ is in $E(G_I, \nu_I)$.  Thus, $\nu_I$ induces a directional cover $\U_I$ of $G_I$ with vertex set $I$ and edge set $E(G_I, \nu_I)$.  Since $E(G_I, \nu_I)$ is precisely the edges of $E(G,\nu)$ among the vertices in $I$, we see that the nerve of the cover $\N(G_I, \U_I)$ is precisely $\N|_I$, the restriction of the nerve of $G$ to the vertices $I$.
\end{proof}

We are now prepared to prove Theorem~\ref{thm:DAG-decomp} (reprinted below for convenience).\\ 

\noindent {\bf Theorem~\ref{thm:DAG-decomp} (DAG decomposition of the nerve)}
For any DAG decomposition $(\W, \T)$ of the nerve $\N$, we have that $G$ is directional with direction $\omega \to \tau$ for $\omega=\pi^{-1}(\W)$ and $ \tau = \pi^{-1}(\T)$.
In particular, 
\[\FP(G) \subseteq \FP(G|_{\tau}),\]
and so for all $\sigma \in \FP(G)$, we have $\pi(\sigma)\subseteq \T$.

\begin{proof}
Let $(\W, \T)$ be a DAG decomposition of the nerve $\N$.  We will show that $G$ is directional with direction $\pi^{-1}(\W) \to \pi^{-1}(\T)$ by inducting on $|\W|$ in the DAG decomposition of $\N$.

The base case of $|\W|=1$ follows immediately from Lemma \ref{lemma:proper_source_nerve} since the first element of $\W$ must be a proper source in $\N$.  For the inductive step, assume the inductive hypothesis holds whenever $|\W|<m$ and consider a DAG decomposition $(\W, \T)$ of $\N$ where $|\W|=m$.  Since $\N|_{\W}$ is a DAG, there is a topological ordering of the vertices  such that the only edges in $\N|_{\W}$ are from lower numbered to higher numbered nodes; WLOG relabel the vertices  of $\W$ as $1, \ldots, m$ according to this ordering.  Let $\W_1 = \W\setminus\{m\}$ and $\T_1 = \T \cup \{m\}$.  It is straightforward to check that $(\W_1, \T_1)$ is also a DAG decomposition of $\N$.  Since $|\W_1|<m$, by the inductive hypothesis, $G$ is directional with 
$$\omega_1 = \pi^{-1}(\W_1) \text{ and } \tau_1 = \pi^{-1}(\T_1) = \pi^{-1}(m) \cup \pi^{-1}(\T).$$
We will show that $G|_{\tau_1}$ is also directional, so that we may apply Lemma~\ref{lemma:dir-iterate} and further refine the directional decomposition of $G$.  Specifically, we will show that $G|_{\tau_1}$ has direction $\omega_2 \to \tau_2$ for $\omega_2 = \pi^{-1}(m)$ and $\tau_2 = \pi^{-1}(\T_1 \setminus \{m\})= \pi^{-1}(\T)$. 

By Lemma~\ref{lemma:restricted-nerve}, the original directional cover of $G$ restricts to a directional cover of $G|_{\tau_1}$ and its nerve is $\N|_{\{m\} \cup \T}$.  Since $m \in \W$ in the original DAG decomposition of $\N$, $m$ is not a sink in $\N$, so it has at least one outgoing edge.  Moreover there are no edges from $\T$ back to $m$ in a DAG decomposition.  Thus $m$ is a proper source in $\N|_{\{m\} \cup \T}$.  Therefore, by Lemma \ref{lemma:proper_source_nerve}, $G|_{\tau_1}$ is directional with direction $\omega_2 \to \tau_2$ for 
$$\omega_2 = \pi^{-1}(m)  \text{ and } \tau_2 = \pi^{-1}(\T).$$

Finally, since $G$ is directional with $\omega_1 = \pi^{-1}(\W \setminus \{m\})$, $\tau_1 =  \pi^{-1}(m) \cup \pi^{-1}(\T)$ and $G|_{\tau_1}$ is directional with $\omega_2 = \pi^{-1}(m)$ and $\tau_2 = \pi^{-1}(\T)$, we see from Lemma~\ref{lemma:dir-iterate}, that $G$ is directional with direction $\omega_1 \cup \omega_2 \to \tau_2$.  Since $\pi^{-1}(\W) = \omega_1 \cup \omega_2$ and $\pi^{-1}(\T) = \tau_2$, we see $G$ is directional with direction $\pi^{-1}(\W) \to \pi^{-1}(\T)$ as desired.
\end{proof}

Next we consider when the nerve $\N$ is itself a DAG so that in the maximal DAG decomposition $\T$ is precisely the sinks of $\N$.  Theorem~\ref{thm:DAG-decomp} guarantees that the directionality of $\N$ pulls back to $G$, but to prove the rest of the nerve theorem conditions, we must appeal to a result characterizing the fixed point supports of \textit{disjoint unions}, proven in \cite{fp-paper}.  The disjoint union of component subgraphs is the graph consisting of those subgraphs with no edges between the components.

\begin{theorem}[\cite{fp-paper}, Theorem 11] Let $G$ be the disjoint union of component subgraphs $G_1,\ldots,G_N$. For any nonempty $\sigma \subseteq V_G$, 
$$\sigma \in \FP(G) \quad \Leftrightarrow \quad \sigma\cap V_{G_i}  \in \FP(G_i) \cup \{\emptyset\} ~~\text{ for all } i \in [N].$$
\label{thm:disjoint-unions} 
\vspace{-.2in}
\end{theorem}

We can now prove Theorem~\ref{thm:DAG} (reprinted below).\\

\noindent{\bf Theorem~\ref{thm:DAG} (DAG nerve)}
Suppose that $\N$ is a DAG, and let $\T = \{\text{sinks of } \N\}$ and $\W=V_\N \setminus \T$.  Then $G$ is directional with direction $\omega \to \tau$ for $\omega=\pi^{-1}(\W)$ and $ \tau = \pi^{-1}(\T)$.
 
 Moreover,
\begin{enumerate}
\item $ \sigma \in \FP(G) \ \Rightarrow \ \pi(\sigma) \in \FP(\N) = \P(\T)\setminus\{\emptyset\}$, where $\P(\T)$ denotes the power set of $\T$. 
\item $ \sigma \in \FP(G) \ \Rightarrow \ \sigma \cap \nu_i \in \FP(G|_{\nu_i}) \cup \{\emptyset\}  \text{ for all } i \in \T$ and  $\sigma \cap \nu_j = \emptyset  \text{ for all } j \in \W$.
\end{enumerate}

\begin{proof}
The fact that  $G$ is directional with direction $\omega \to \tau$ for $\omega=\pi^{-1}(\W)$ and $ \tau = \pi^{-1}(\T)$ follows from Theorem~\ref{thm:DAG-decomp} since the given choice of $(\W,\T)$ is the maximal DAG decomposition of $\N$.  As a consequence of this, we have $\FP(G) \subseteq \FP(G|_{\pi^{-1}(\T)})$, and so we turn our attention to $G|_{\pi^{-1}(\T)}$ to understand $\FP(G)$.

Observe that since $\T = \{\text{sinks of } \N\}$, there are no edges between the vertices  in $\T$, and so $\N|_\T$ is an independent set.  Thus,  there are no edges in $G$ between the components $\nu_i$ for $i \in \T$, and so $G|_{\pi^{-1}(\T)}$ is a disjoint union of the component subgraphs $G|_{\nu_i}$.  Applying Theorem~\ref{thm:disjoint-unions}, we see that $\sigma \in \FP(G|_{\pi^{-1}(\T)})$ precisely when $\sigma \cap \nu_i \in \FP(G|_{\nu_i}) \cup \{\emptyset\} $ for all $i \in \T$.  And since $\FP(G) \subseteq \FP(G|_{\pi^{-1}(\T)})$ by the directionality of $G$, part (2) of the theorem statement follows immediately.  

For part (1), observe that since $G$ is directional, $\sigma \in \FP(G)$ implies that $\pi(\sigma) \subseteq \T$.
Since $\N$ is a DAG, by Lemma~\ref{lemma:FP-DAG}, $\FP(\N) = \P(\T)$, and so every subset of $\T$ is an element of $\FP(\N)$.  Thus, $ \sigma \in \FP(G) \ \Rightarrow \ \pi(\sigma) \in \FP(\N)$ as desired.
\end{proof}

\section{Some extensions and applications} \label{sec:applications}

We now turn our attention to some examples that illustrate the power of our nerve theorems. Going back to Figure~\ref{fig:clique-chain}A,B of the Introduction, we see that this graph and its nerve satisfy the hypotheses of Theorem~\ref{thm:DAG-decomp}  and Theorem~\ref{thm:DAG}. The nerve (panel B) is a simple path that has a maximal DAG decomposition with $\T = \{10\}$ ($10$ is the unique sink node). Theorem~\ref{thm:DAG-decomp} thus predicts that $\FP(G) \subseteq \FP(G|_{\nu_{10}})$, since $\nu_{10} = \pi^{-1}(\{10\}).$ 
In fact, $\FP(G) = \{\nu_{10}\}$, so the network has a unique fixed point supported on $\nu_{10}$. 
Moreover, this fixed point is stable because it corresponds to a clique (see \cite{fp-paper, stable-fp-paper}).
As seen in panel D, the dynamics do indeed converge to this stable fixed point. Furthermore, for solutions with initial conditions supported on the first clique ($G|_{\nu_1}$), we see that the transient dynamics activate all cliques in the chain, in sequence, following the path of the nerve.

In the remainder of this section, we will discuss additional examples of networks whose graphs and nerves satisfy the hypotheses of one or more of our nerve theorems: Theorem~\ref{thm:DAG-decomp}, Theorem~\ref{thm:DAG}, and Theorem~\ref{thm:cycle-nerve}.  Just as in Figure~\ref{fig:clique-chain}, we will see that the nerve not only predicts the fixed points and asymptotic dynamics, but also provides insight into the transient dynamics of the network. In the second subsection, we will see that even when we violate a key condition of directional covers, the nerve of a network covered by directional graphs can still provide accurate predictions of the dynamics.

\subsection{Iterating and combining DAG decomposition and cycle nerve theorems}

Recall that for any partition $\{\nu_1,\ldots,\nu_n\}$ of the vertices of $G$, there is an associated quotient map $\pi:V_G\rightarrow V_\N=[n]$ defined by $\pi(\nu_i) = \lbrace i\rbrace$ for each $i \in [n]$. We saw in Section~\ref{sec:dir-covers} that if such a partition induces a directional cover, and the nerve $\N$ has a DAG decomposition $(\W,\T)$, then the fixed points of the corresponding CTLN are confined to the non-DAG part $\T$.  In other words, Theorem \ref{thm:DAG-decomp} tells us that
$$\FP(G) \subseteq \FP(G|_\tau),$$
where $\tau = \pi^{-1}(\T)$. Now if we consider the restricted graph, $G' = G|_\tau$, we can potentially iterate this process by finding a new directional cover, with nerve $\N',$ DAG decomposition $(\W',\T')$, and quotient map $\pi'$. This would enable us to further restrict the fixed points of the original network to
$$\FP(G) \subseteq \FP(G') \subseteq \FP(G''),$$
where $G'' = G|_{\tau'}$, and $\tau' = \pi'^{-1}(\T') \subset \tau$. Note that $G''$ is the original graph restricted to an even smaller subset of vertices. This kind of iteration may enable us to get more power from our nerve theorems, by further constraining $\FP(G)$.

\begin{figure}[!ht]
  \centering
  \includegraphics[width=6.5in]{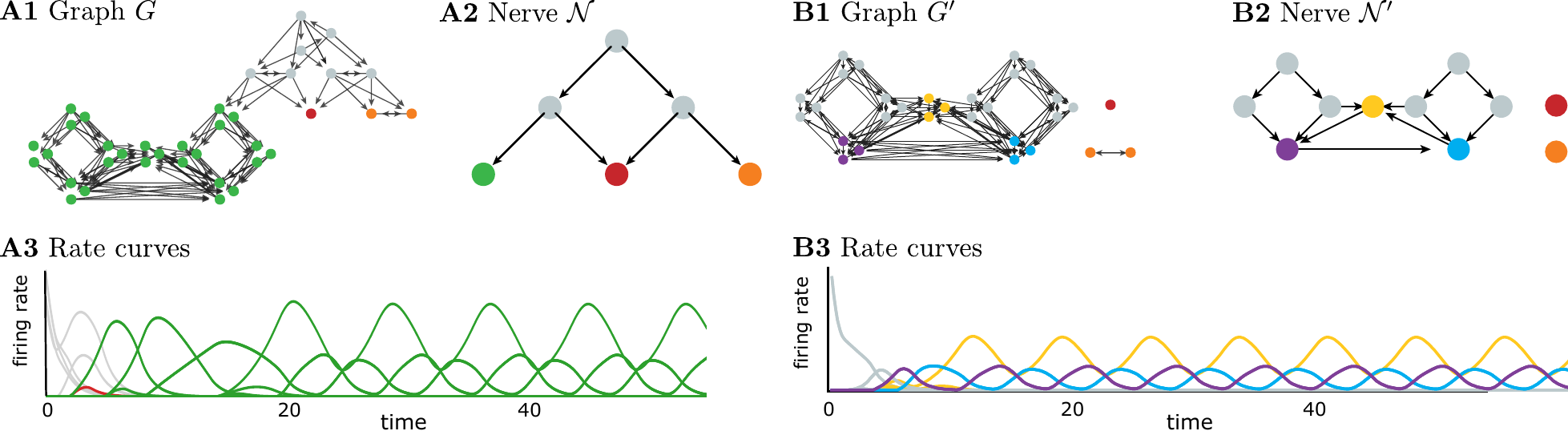}
  \caption{Iterative application of nerve theorems.} 
  \label{fig:sim_recursion_of_nerve}
\end{figure}

Figure~\ref{fig:sim_recursion_of_nerve} shows an example of a graph (A1) whose nerve (A2) is a DAG. Therefore, we can apply Theorem~\ref{thm:DAG} to conclude that the fixed point supports of $\FP(G)$ must be unions of fixed points of $G$ restricted to the green, red, or orange components. Moreover, the activity flows towards the subnetwork corresponding to the sinks in the nerve. Indeed, for a given initial condition supported on the top gray nodes, we see the solution converge to a limit cycle supported only on green nodes (A3). 

The green component, however, is itself a complex graph. Thus, we may consider the subgraph $G'$ of $G$ corresponding to the non-DAG part of $\N$.  Figure~\ref{fig:sim_recursion_of_nerve} (right) shows $G'$ (B1) with nodes colored according to a DAG decomposition of its own nerve $\N'$ (B2). (Note that $G'$ is a disjoint union of three graphs, and the nerve $\N'$ is the disjoint union of nerves for each connected component of $G'$.) Applying Theorem \ref{thm:DAG-decomp} allows us to restrict the fixed points of $G$ even further, to the vertices that map to the colored nodes in B2. We see this reflected in the dynamics as well. Panel B3 shows the same limit cycle as before, only now it's clear that the green curves in A3 correspond only to the yellow, purple, and blue neurons in B1.

\begin{figure}[!ht]
  \centering
  \includegraphics[width=4.5in]{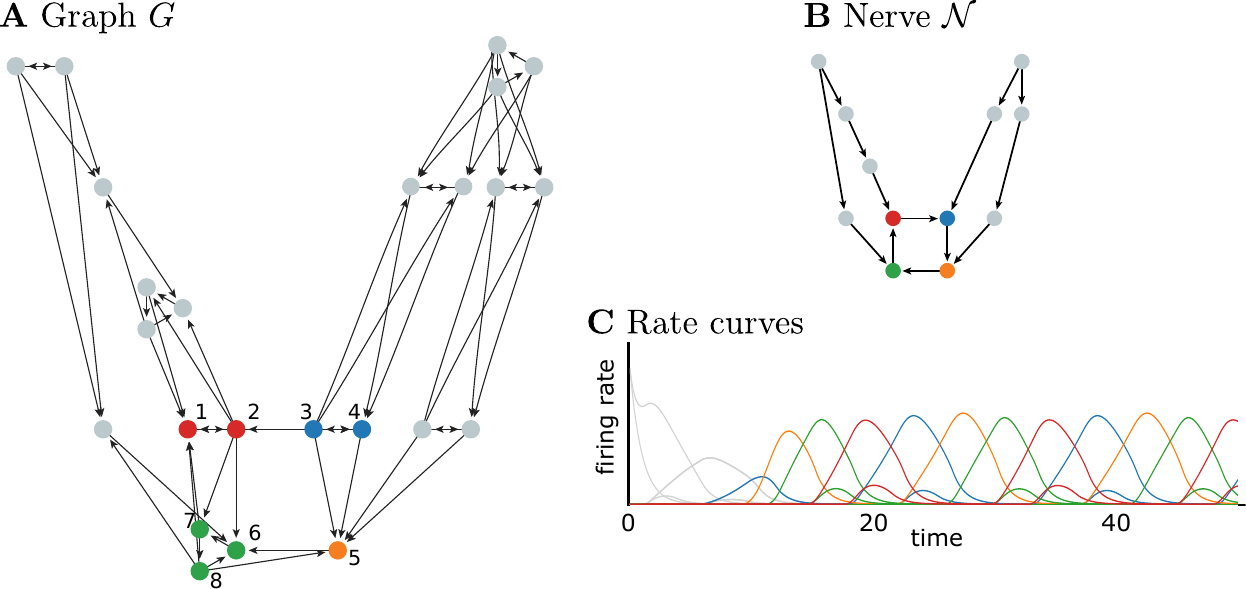}
  \vspace{.05in}
  \caption{Application of multiple nerve theorems. (A) A graph $G$ for a CTLN. (B) A nerve $\N$ of $G$ with 12 vertices. Each of the gray vertices in $\N$ corresponds to a cluster of gray nodes in (A), while the colored vertices in $\N$ correspond to the vertices with matching colors in $G$. (C) A solution to a CTLN with graph $G$ and initial conditions supported on the top-most gray nodes. The activity flows down the network and converges to a limit cycle involving only the colored nodes.}
  \label{fig:sim_complicated_graph_and_nerve}
\end{figure}

We can also combine the DAG nerve theorems with the cycle nerve theorem, Theorem~\ref{thm:cycle-nerve}. Figure~\ref{fig:sim_complicated_graph_and_nerve}A depicts the graph of a complex network whose nodes are grouped according to a partition with 12 components (8 in gray, 4 in color). In panel B we see the nerve of the induced directional cover, with the color of each node matching those in the original graph. This nerve has a DAG decomposition with the eight gray nodes in the DAG part, $\W$, and the four colored nodes in the non-DAG part, $\T$. Theorem~\ref{thm:DAG-decomp} thus tells us that all fixed points of the original graph $G$ must be contained in $\tau = \pi^{-1}(\T)$, which is the set of colored nodes in panel A. Indeed, $\FP(G) \subseteq \FP(G|_\tau)$ for this graph. 

Note that we can also apply another nerve theorem to $G|_\tau$ to say something stronger about $\FP(G)$. Because $G|_\tau$ has a nerve that is a cycle, Theorem~\ref{thm:cycle-nerve} tells us that any fixed point of $G$ must intersect each of the components $\pi^{-1}(i)$ for $i$ in the cycle. That is, fixed points of $G$ must contain at least one vertex from each of the four colors (red, blue, green, orange) shown in panel A. This is in fact the case, as the CTLN for $G$ has $\FP(G) = \{14567\}$. Moreover, we see that even if we choose initial conditions supported only on the gray vertices, the dynamics will converge to a part of the state space where the gray neurons are off and at least one neuron of each color is active (see Figure~\ref{fig:sim_complicated_graph_and_nerve}C).

\subsection{Extensions beyond directional covers}

In Definition \ref{def:directional_cover} of (rigid) directional covers, we required fairly stringent conditions that enabled us to prove strong results about the fixed points of a graph in terms of the fixed points of its nerve. Here we consider some examples of ``weak'' directional covers, where the component graphs are all directional but one of the conditions in Definition \ref{def:directional_cover} is violated. Nevertheless, we find that the nerve provides a remarkably accurate prediction of the network dynamics.

\begin{figure}[!ht]
  \centering
  \includegraphics[width=6.5in]{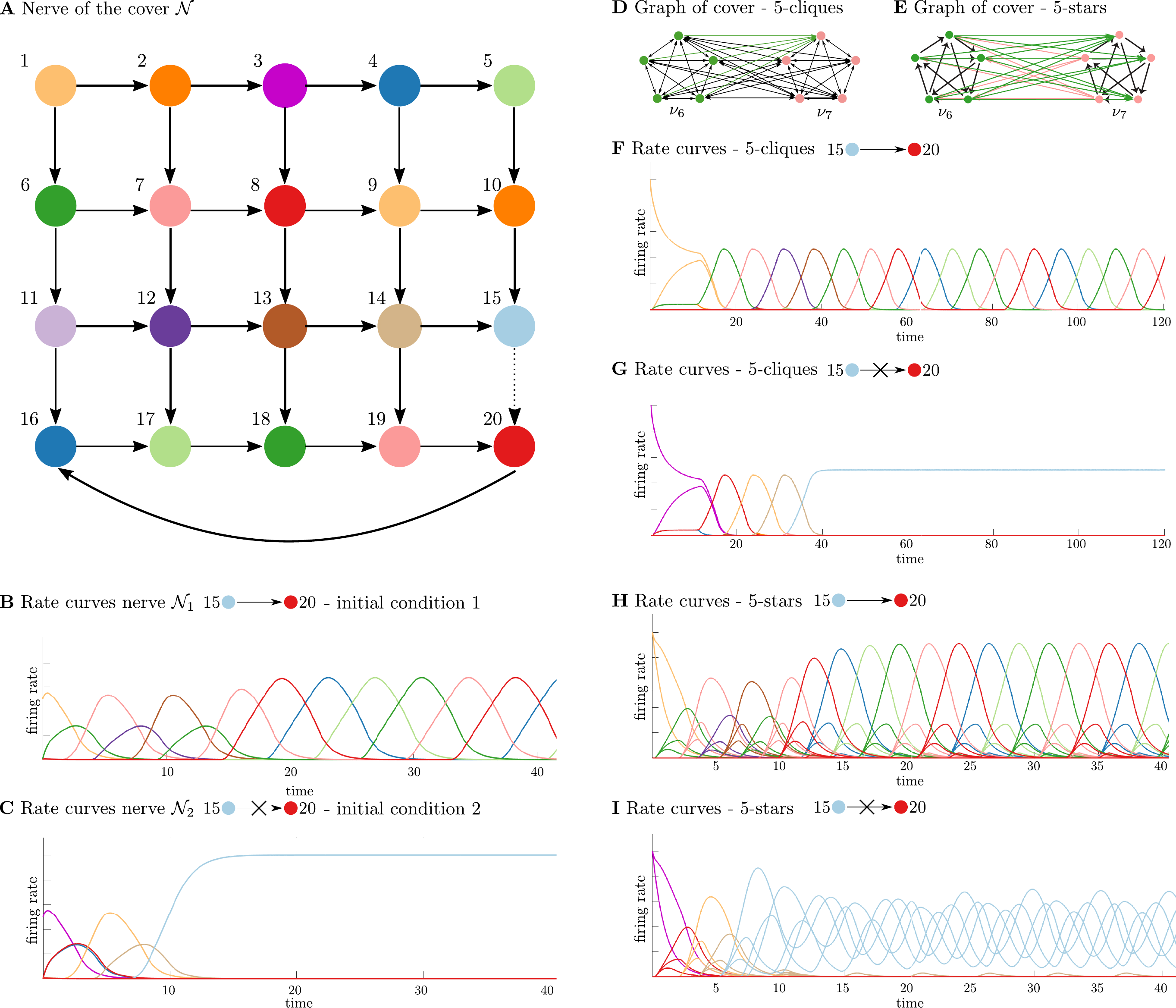}
  \caption{A grid-like nerve shapes global sequential dynamics irrespective of the component graphs. (A) Two versions of the nerve, with and without the $15 \to 20$ edge. (B-C) Solutions for CTLNs where the graph is nerve 1 or nerve 2, respectively. (D) $5$-clique component graphs and a corresponding directional graph for an edge in the cover. (E) $5$-star component graphs and their directional graph. (F-I) Solutions for associated CTLNs for all four combinations of nerves and component graphs.}
  \label{fig:nerve-grid}
\end{figure}

Our starting point for these examples is a pair of ``nerve'' graphs with a grid-like structure, shown in Figure~\ref{fig:nerve-grid}A. Each graph is a finite lattice with directed edges moving down and to the right across the grid, and an additional edge, $20 \to 16$, completing a cycle at the bottom. We can think of these graphs as a pair of nerves, $\N_1$ and $\N_2$, for larger networks obtained by inserting directional graphs along the edges. In this case, a vertex $i$ corresponds to a component $\nu_i$, and an edge $i \to j$ corresponds to a directional graph $G|_{\nu_i \cup \nu_j}$ with direction $\nu_i \to \nu_j$. The only difference is that the first nerve, $\N_1$, includes the $15 \to 20$ edge; while the second nerve, $\N_2$, does not (see dotted line in Figure~\ref{fig:nerve-grid}A). 

The nerves enable one to make concrete predictions about the network dynamics. Specifically, we can consider CTLNs where the graph $G$ is chosen to be $\N_1$ or $\N_2$. Figure~\ref{fig:nerve-grid}B shows a solution for a CTLN with $G = \N_1$, where the activity is initialized with $x_1(0) = 0.5$ and $x_i(0) = 0$ for $i>1$ (we refer to this as initial condition 1). Note that the transient activity follows a hopscotch trajectory $1 \to (2,6) \to 7 \to (8,12) \to 13 \to (14,18) \to 19$, where the pairs $ (2,6), (8,12),$ and $(14,18)$ fire synchronously. Once the activity reaches the bottom row $\T$, the dynamics converge to a limit cycle that follows the cycle $ 19 \to 20 \to 16 \to 17 \to 18 \to 19$ in the graph. If we choose the same initial condition for $G = \N_2$, we obtain exactly the same result. This is because the activity never reaches node $15$, and so the cut edge is not ``seen.'' Figure~\ref{fig:nerve-grid}C shows the solution when we initialize instead at $x_3(0) = 0.5$, and $x_i(0) = 0$ for $i\neq 3$ (initial condition 2). In this case, the activity follows the hopscotch trajectory $3 \to (4,8) \to 9 \to (10,14) \to 15$ that ends in a stable fixed point at $15$. This is precisely what we expect since node $15$ is a sink. These basic features of the dynamics for a CTLN defined directly on the nerve can be taken as a prediction for the dynamics of any network $G$ that has a directional cover with $\N_1$ or $\N_2$ as its nerve.

If we consider a graph $G$ for which $\N_1$ or $\N_2$ is the nerve of a directional cover, satisfying all the conditions of Definition \ref{def:directional_cover}, then we have additional knowledge about the fixed points of $G$. 
The first nerve, $\N_1$, has a maximal DAG decomposition $(\W,\T)$ with nodes 1-15 in $\W$ and nodes 16-20 in $\T$. Theorem~\ref{thm:DAG-decomp} thus predicts that all fixed points of $G$ are supported in $\tau = \pi^{-1}(\T)$. Moreover, since $\N|_\T$ forms a cycle, applying Theorem~\ref{thm:cycle-nerve} we expect all fixed points to intersect each component of $\T$. In the second nerve, $\N_2$, $15 \not\to 20$ and thus node 15 is a sink. Any DAG decomposition for $\N_2$ must therefore include node 15 in $\T$. Applying Theorem~\ref{thm:DAG-decomp}, we expect fixed point(s) corresponding to the cycle, as before, as well as fixed point(s) supported in $\nu_{15} = \pi^{-1}(15)$. We may also have fixed points whose supports are unions of those from the cycle and $\nu_{15}$. These observations are all independent of the choice of components $\nu_i$ of $G$ and the subgraphs $G|_{\nu_i \cup \nu_j}$, so long as the nerve corresponds to a directional cover in accordance with Definition \ref{def:directional_cover}. 

What happens if a cover by directional graphs violates one of the conditions of Definition \ref{def:directional_cover}? Figure~\ref{fig:nerve-grid}D shows a directional graph that can be inserted along the edges of $\N_1$ and $\N_2$. The components are $5$-cliques, and the directional graphs are the same as in Figure~\ref{fig:clique-chain}C. Inserting these into the grid-like nerves, however, yields many nodes with splitting overlaps (unlike for the nerve in Figure~\ref{fig:clique-chain}B). This means backwards edges within the directional graphs violate the ``splitting'' condition $2$ of Definition \ref{def:directional_cover}, and the resulting network does not have a directional cover. Moreover, the splitting condition was essential to our nerve theorems: the fixed points of a network $G$ obtained by inserting the Figure~\ref{fig:clique-chain}C graph into either $\N_1$ or $\N_2$ do not satisfy the constraints given by Theorem~\ref{thm:DAG-decomp}. In fact, there are numerous fixed points supported outside $G|_\tau$ for $\tau = \pi^{-1}(\T)$.

As another example, inserting the directional graph in Figure~\ref{fig:nerve-grid}E as $G|_{\nu_i \cup \nu_j}$ along the edges of $\N_1$ or $\N_2$ also produces a cover by directional graphs that violates the splitting condition.  Here, each component is a cyclically symmetric oriented graph on five vertices, called the $5$-star,\footnote{Each vertex $k$ in the $5$-star has two outgoing edges: $k \to k+1$ and $k \to k+2$ (indexing mod $5$).} and the directional graphs $G_{ij} = G|_{\nu_i\cup\nu_j}$ are chosen to have forward edges from $\nu_i$ onto three of the nodes in $\nu_j$, and backwards edges from the remaining two nodes in $\nu_j$ to all five nodes in $\nu_i$. Again, our nerve theorems fail to predict the fixed point structure of the larger network.

Nevertheless, we find that the dynamics of these networks whose graph covers violate the splitting condition are well predicted by their nerves. Figure~\ref{fig:nerve-grid}F-I display the dynamics of the networks obtained from each of the four combinations: $5$-clique components with nerve $\N_1$ (panel F), $5$-clique components with nerve $\N_2$ (panel G), $5$-star components with nerve $\N_1$ (panel H), and $5$-star components with nerve $\N_2$ (panel I). In each case, the nerve dictates the global structure of the dynamics as activity flows from one component to another. 
Figure~\ref{fig:nerve-grid}F shows the solution for a CTLN with $5$-clique components and initial conditions supported in $\nu_1$. As predicted, the asymptotic behavior is of a limit cycle following the cycle $16 \to 17 \to 18 \to 19 \to 20 \to 16$ in the nerve.  Figure~\ref{fig:nerve-grid}G shows the activity with $5$-clique components inserted into nerve $\N_2$, with initial conditions leading the activity to converge to a stable fixed point, as in Figure~\ref{fig:nerve-grid}C.

Inserting $5$-star graphs in $\N_1$ and $\N_2$, instead of $5$-cliques, also produces asymptotic behavior that settles into a repeating sequence (Figure~\ref{fig:nerve-grid}H) or a localized attractor (Figure~\ref{fig:nerve-grid}I).
The structure of the inserted graphs, however, does affect the local dynamics within each component of the nerve. In Figure~\ref{fig:nerve-grid}F the transient dynamics are considerably more regular than in Figure~\ref{fig:nerve-grid}B, and the neurons within each component fire synchronously. In contrast, in Figure~\ref{fig:nerve-grid}H the transient dynamics are irregular like in Figure~\ref{fig:nerve-grid}B, and the neurons within each component do not fire synchronously.
In each case, we see global aspects of the dynamics being dictated by the nerve, while local dynamics are affected by differences in the component graphs $G|_{\nu_i}$. For example, in Figure~\ref{fig:nerve-grid}G the activity converges to a stable fixed point corresponding to the $5$-clique supported on $\nu_{15}$; but in Figure~\ref{fig:nerve-grid}I, the network does not converge to a fixed point because the $5$-star does not support a stable fixed point. Instead, the activity settles into a limit cycle typical of $5$-star CTLNs, with activity confined almost entirely to the neurons in $\nu_{15}$. 

Taken together, these examples suggest that nerves can be predictive of network dynamics for a broader class of directional covers, including networks where our current set of nerve theorems do not apply. It is an open question how to formalize these observations into new nerve theorems that reflect the predictions on the dynamics.

\section{Conclusion} \label{sec:conclusion}

In this work, we investigated how the global structure of a network, as captured by the nerve of a directional cover, reflects the underlying dynamics. By replacing directional subgraphs with single edges, the nerve provides a significant dimensionality reduction of a network. Moreover, this reduced network is meaningful: in simulations, we have seen that the dynamics of a CTLN with a directional cover closely follows the dynamics of its nerve.  

Although the observed relationship between the dynamics of a network and its nerve is so far heuristic, we have proven a number of theorems directly connecting the fixed points of a larger network $G$ to those of its nerve $\N$.  Specifically, we showed that whenever the nerve has a \emph{DAG decomposition} $(\W, \T)$, the nerve is directional with direction $\W \to \T,$ and this guarantees that the larger network $G$ is also directional.  In particular, the fixed points of the larger network are confined to live in the pull-back $\pi^{-1}(\T)$ (Theorem~\ref{thm:DAG-decomp}).  In the special case where the nerve is a DAG, we have a tighter connection between $\FP(G)$ and $\FP(\N)$.  Theorem~\ref{thm:DAG} shows that every fixed point support of $G$ projects to a fixed point of $\N$. In other words,
$$ \sigma \in \FP(G) \ \Rightarrow \ \pi(\sigma) \in \FP(\N).$$
Moreover, every $\sigma \in \FP(G)$ is a union of fixed point supports of the component subgraphs $G|_{\vu_i}$.  
Theorem~\ref{thm:cycle-nerve} gives similar constraints on $\FP(G)$ whenever the nerve is a cycle.

Due to the close relationship between fixed points and attractors in CTLNs, understanding the fixed point structure and how this is shaped by network architecture is an important step towards understanding how network connectivity shapes dynamics.  The machinery developed here thus provides a useful framework for dimensionality reduction in the analysis of large networks.  Moreover, it provides insight into how to engineer complex networks with desired dynamic properties from smaller building block components.

\begin{acknowledgement}
This research is a product of one of the working groups at the Workshop for Women in Computational Topology (WinCompTop) in Canberra, Australia (1-5 July 2019).  We thank the organizers of this workshop and the funding from NSF award CCF-1841455, the Mathematical Sciences Institute at ANU, the Australian Mathematical Sciences Institute (AMSI), and Association for Women in Mathematics that supported participants' travel.  
We thank Caitlyn Parmelee for fruitful discussions that helped set the foundation for this work.  We would also like to thank Joan Licata for valuable conversations at the WinCompTop workshop.  CC and KM acknowledge funding from NIH R01 EB022862, NIH R01 NS120581, NSF DMS-1951165, and NSF DMS-1951599.
\end{acknowledgement}







\end{document}